\documentclass[comment, prd,twocolumn,
nofootinbib, preprintnumbers, superscriptaddress]{revtex4}

\usepackage{amsmath,amssymb,bm,slashed,braket}
\usepackage{graphicx}
\usepackage{epstopdf}
\usepackage{float}
\usepackage{xcolor}
\usepackage{multirow}
\usepackage{booktabs}   
\usepackage{pifont}     
\usepackage{dsfont}
\usepackage{placeins}
\usepackage[colorlinks=true,
            linkcolor=blue,
            urlcolor=blue,
            citecolor=green,          
            bookmarks=true,
            bookmarksnumbered=true,
            breaklinks=true,
            pdfpagemode=Fullscreen,
            pdfstartview=FitBH]{hyperref}
\usepackage[normalem]{ulem}
\allowdisplaybreaks[4]
\usepackage[capitalise]{cleveref}

\newcommand{\keV}{\rm keV}

\begin{document}

\title{Interpretation of the LZ Event via Inelastic ALP-Dark Photon Transition with Photon Exchange}
	

\author{Quanyu Wang}
\email[]{wangquanyu@htu.edu.cn}
\affiliation{School of Physics, Henan Normal University, Xinxiang 453007, P. R. China}

\author{Xiangwei Yin}
\email[]{yinxiangwei@htu.edu.cn (corresponding author)}
\affiliation{School of Physics, Henan Normal University, Xinxiang 453007, P. R. China}

\author{Tianjun Li}
\email[]{tli@itp.ac.cn}
\affiliation{School of Physics, Henan Normal University, Xinxiang 453007, P. R. China}

\begin{abstract}
	
The LUX-ZEPLIN (LZ) experiment has reported a nuclear recoil event at 
$248\pm23$~keV with a global significance of about $2.6\sigma$. We for the first time explain it via the inelastic process with photon mediator. We consider a dark matter (DM) axion-like particle (ALP) $a$ converts into a heavier dark
photon $\gamma'$ via the nuclear electromagnetic current, and study the photon-mediated endothermic transition
$aA\rightarrow\gamma'A$. We derive the kinematics and
cross sections for non-relativistic halo and boosted ALPs. A three-bin Poisson
fit to the LZ data includes the xenon isotopes, charge form factors, detector
efficiency, and energy resolution. Halo DM, which has mass  $m\simeq1.2$~TeV, mass splitting
$\delta\simeq346$~keV, and near the upper coupling bound, reproduces the event with
about $61\%$ in the middle bin. Boosted DM with $v_{0}=0.05\,c$ and
$\Phi_{0}=10^{5}~\mathrm{cm^{-2}\,s^{-1}}$ requires only $m\simeq5.74$~GeV and
$\delta\simeq6.66$~MeV to give about $62\%$. We find that a broader Gaussian spread in the incident energy has a significant effect on the fitted masses.

\end{abstract}

\maketitle

\section{Introduction}
Dark matter (DM) is inferred from several gravitational observations: galaxy rotation curves indicate unseen mass beyond the visible matter~\cite{Rubin:1970zza}. Gravitational lensing of the Bullet Cluster shows mass separated from its hot gas~\cite{Clowe:2006eq}. Cosmic microwave background data also support a cold DM component in the universe~\cite{Planck:2018vyg}. However, the particle nature of DM remains unknown. 

Electron-recoil experiments search for ionization or excitation signals produced when DM interacts with electrons in atoms or crystals, making them particularly well suited to sub-GeV DM particle searches~\cite{Essig:2015cda,Essig:2017kqs,Crisler:2018gci,SENSEI:2023zdf}. Nuclear-recoil experiments detect the recoil energy imparted to nuclei by DM scattering and are a leading approach to searching for DM with masses of GeV or higher~\cite{Lewin:1995rx,LZ:2022lsv,XENON:2023cxc,PandaX:2022aac,SuperCDMS:2024yiv}. Recently, the LUX-ZEPLIN (LZ) experiment extended its recoil-energy search to approximately $270\,\mathrm{keV}$ and reported a candidate event at $248\pm23\,\mathrm{(stat)}\pm23\,\mathrm{(sys)}\,\mathrm{keV}$ in an exposure of $2.84$ tonne-years~\cite{LZ:2026axp}. The excess has a maximum local significance of $3.4\sigma$, corresponding to a global significance of $2.6\sigma$ after accounting for the look-elsewhere effect. Although this falls short of discovery, it motivates investigating whether DM scattering can produce high-energy recoils without a corresponding low-energy excess.

For standard contact spin-independent elastic scattering with fixed DM and halo parameters, the recoil spectrum falls with increasing energy. Thus, increasing the rate to explain high-energy recoils inevitably increases the predicted low-energy contribution as well~\cite{Lewin:1995rx,PDG2025}. Exothermic inelastic scattering makes heavier DM state scattering into a lighter one, releasing mass-splitting energy that can contribute to the nuclear recoil~\cite{Graham:2010ca,Bell:2021zkr}. This can yield detectable signals even for slow DM, but the rate depends on the local abundance of the heavier state and the specific model; moreover, the released energy is not necessarily transferred entirely to the nucleus. Endothermic inelastic scattering may provide a more suitable explanation. Transition to a heavier DM state requires a velocity threshold and can suppress low energy recoils~\cite{Tucker-Smith:2001myb,XENON:2022avm}.

Axion-like particle (ALP) are hypothetical pseudoscalar particles whose masses and coupling strengths are not linked as they are for the QCD axion. One possible way to detect ALPs is through their couplings to photons~\cite{Jaeckel:2010ni,Ringwald:2014vqa}. A dark photon is a hypothetical gauge boson of a dark-sector $U(1)$ symmetry that may interact weakly with ordinary matter through kinetic mixing with the Standard Model photon or hypercharge field~\cite{Jaeckel:2010ni,Holdom:1985ag}. Dark photon-mediated inelastic DM scattering has been constrained using nuclear recoil data from XENON1T, CRESST-III, CDEX-1B, and DarkSide-50, with later PandaX-4T analyses extending the search to low-mass DM via the Migdal effect~\cite{He:2020sat,PandaX:2023xgl}. ALP-mediated elastic and inelastic DM–nucleus scattering has also been explored for xenon-based direct detection~\cite{Bae:2023ago,Beenakker:2025mhf}. Several recent works interpret LZ experiment through ALP or dark-photon interactions. ALP models explain the candidate through either inelastic scattering, which suppresses low-energy recoils, or elastic scattering with a harder spectrum~\cite{Yuan:2026djt,Unwin:2026rdp}. Dark-photon models invoke endothermic or exothermic scattering, or treat the dark photon itself as DM~\cite{Yamashita:2026ump,deLima:2026shq,Zhu:2026dag}.
Other DM–nucleus interaction mechanisms have also been proposed to explain this high-recoil signal, such as Higgsino, sneutrino, and bino-like models~\cite{Yin:2026jnn,Fan:2026kxx,Freese:2026sga,Du:2026guj,Lian:2026hpm,Roy:2026wzb}, as well
as scalar or fermion DM models~\cite{Wang:2026ytg,Bandyopadhyay:2026gjw,Borah:2026zwf}. Boosted DM can also offer an explanation, allowing the DM particle to be much lighter~\cite{Liang:2026coz,Kannike:2026qyl,Alhazmi:2026efz,Heikinheimo:2026kwp,Chauhan:2026udz}.

In this work, we for the first time investigate photon-mediated ALP-dark photon transitions as a possible explanation for the high-energy recoil event reported by LZ. We consider the endothermic inelastic process $aA \to \gamma' A$, in which the ALP $a$ converts into a heavier dark-photon state $ \gamma'$ via the nuclear electromagnetic current. Starting from the effective ALP-dark photon interaction, we evaluate both the
non-relativistic and the boosted-DM transitions, as well as derive the corresponding kinematics and cross sections. A simplified three-bin Poisson likelihood is used to fit the LZ data, incorporating the natural abundance of xenon isotopes together with the detector efficiency and energy resolution.

For non-relativistic DM, we scan the mass $m$, the mass splitting $\delta$ and coupling $G_{a\gamma\gamma'}$. The scattering rate is integrated over the laboratory-frame Standard Halo Model (SHM) velocity distribution~\cite{Drukier:1986tm}, with the speed-dependent flux included, to compute the expected number of events. We find a best-fit benchmark with $m\simeq1.2~{\rm TeV}$, $\delta\simeq346~{\rm keV}$ with the coupling $G_{a\gamma\gamma'}=0.1\mathrm{GeV}^{-1}$ near the upper bound. This benchmark probes the high-speed tail and provides one accepted event with about $61\%$ in the middle bin.

For boosted DM, we consider velocities above the halo cutoff and take the initial DM flux as an input, leaving the boost mechanism unspecified. We examine both monochromatic and Gaussian-distributed incident DM, scanning over the DM mass $m$ and the mass splitting $\delta$ at fixed coupling. A benchmark point with a comparable fit is found at $v_{0}=0.05\,c$ and $\Phi_{0}=10^{5}\,\mathrm{cm}^{-2}\,\mathrm{s}^{-1}$, with
$m\simeq5.74\,\mathrm{GeV}$ and $\delta\simeq6.66\,\mathrm{MeV}$, predicting about $62\%$ of the events in the middle bin. We also evaluate the impact of the Gaussian spread in the initial state on the fitted masses. 

This paper is organized as follows. In Sec.~\ref{sec:theory}, we introduce the interactions and kinematics for both non-relativistic and boosted DM. Sec.~\ref{sec:statistic} describes the calculation of event-rate and statistical analysis. In Sec.~\ref{Sec:result} we present benchmarks and parameter regions for both cases. We conclude in Sec.~\ref{Sec:conclude}.

\section{ALP portal and endothermic nuclear scattering}
\label{sec:theory}
\subsection{Interaction and conventions}
We consider a massive ALP $a$ and a massive
dark photon $\gamma^{\prime}$, coupled through the dark axion
portal \cite{Kaneta:2016wvf,Kaneta:2017wfh}.
The relevant low-energy Lagrangian is
\begin{align}
\mathcal{L}\supset{}&
\frac{1}{2}\partial_\mu a\,\partial^\mu a
-\frac{1}{2}m_a^2a^2
-\frac{1}{4}F_{\mu\nu}F^{\mu\nu}
-\frac{1}{4}F_{\mu\nu}^{\prime}F^{\prime\mu\nu}
\nonumber\\
&+\frac{1}{2}m_{\gamma^{\prime}}^2
A_\mu^{\prime}A^{\prime\mu}
+\frac{G_{a\gamma\gamma^{\prime}}}{2}\,
aF_{\mu\nu}\widetilde F^{\prime\mu\nu},
\label{eq:lagrangian}
\end{align}
where $a$ is a real pseudoscalar field and $A_\mu^{\prime}$
is the vector field describing the dark photon.
The electromagnetic and dark-photon field strengths are
$F_{\mu\nu}=\partial_\mu A_\nu-\partial_\nu A_\mu$ and
$F_{\mu\nu}^{\prime}
=\partial_\mu A_\nu^{\prime}-\partial_\nu A_\mu^{\prime}$,
respectively, with $A_\mu$ denoting the photon field.
The dual field strength is defined as
$\widetilde F^{\prime\mu\nu}
=\epsilon^{\mu\nu\rho\sigma}F_{\rho\sigma}^{\prime}/2$.
The masses of the two dark-sector particles are $m_a$ and
$m_{\gamma^{\prime}}$, and
$G_{a\gamma\gamma^{\prime}}$ is the mixed portal coupling.
The interaction of $G_{a\gamma\gamma^{\prime}}$ has been investigated via concrete model~\cite{Kaneta:2016wvf,Kaneta:2017wfh}.
Here, we do not specify an ultraviolet completion and instead treat the dark-sector masses and the coupling as free parameters.

\begin{figure}[htbp]
    \centering
    \includegraphics[width=0.6\columnwidth]{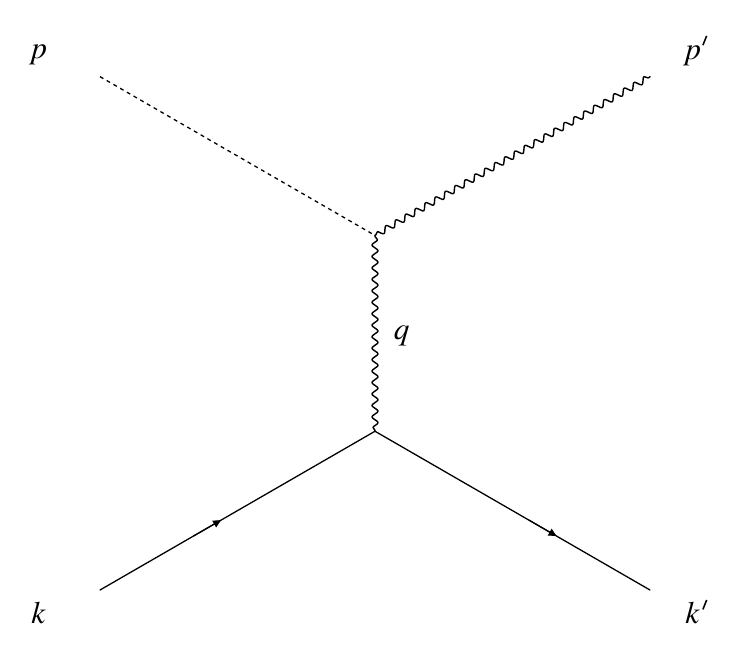}
    \caption{Feynman diagram of axion dark photon transition. The interaction with nucleus is by virtual photon.}
    \label{fig_Feyd}
\end{figure}

We consider photon-mediated inelastic scattering in which an incident ALP converts into a dark photon upon scattering
from a nucleus,
\begin{equation}
a(p)+A(k)
\longrightarrow
\gamma^{\prime}(p^{\prime})+A(k^{\prime}),
\label{eq:scattering_process}
\end{equation}
where $A$ denotes the target nucleus, with mass $M_A$ and
electric charge $Ze$.
The nucleus remains in the same internal state, so the
inelasticity arises entirely from the dark-sector transition.
The corresponding Feynman diagram is shown in
FIG.~\ref{fig_Feyd}. We adopt the following mass hierarchy
\begin{equation}
m_a=m,
\qquad
m_{\gamma^{\prime}}=m+\delta,
\qquad
\delta>0,
\label{eq:mass_convention}
\end{equation}
where $m$ is the incident ALP mass and $\delta$ is the positive
mass splitting.
The reaction is therefore endothermic: part of the initial
kinetic energy is required to produce the heavier dark photon.
We assume that the lighter ALP is stable and constitutes the DM candidate.
The parameters $m$, $\delta$, and $G_{a\gamma\gamma^{\prime}}$ are treated as independent.

In the laboratory frame, the initial nucleus is at rest,
$k^\mu=(M_A,\boldsymbol{0})$.
We define the four-momentum transfer and the nuclear recoil
energy as
\begin{equation}
q^\mu=k^{\prime\mu}-k^\mu=p^\mu-p^{\prime\mu},
\qquad
E_R=k^{\prime 0}-M_A.
\label{eq:momentum_transfer}
\end{equation}
Writing $t=q^2$ and $Q=|\boldsymbol q|$, the nuclear
on-shell conditions imply
\begin{equation}
t=-2M_AE_R,
\qquad
Q^2=2M_AE_R+E_R^2.
\label{eq:transfer_recoil}
\end{equation}
Hence, $Q\simeq\sqrt{2M_AE_R}$ in the non-relativistic
nuclear-recoil limit.

\subsection{Non-relativistic kinematics and coherent charge cross section}
The incident ALP has
laboratory-frame speed $v$ and kinetic energy $mv^2/2$ at
leading order. In the non-relativistic limit
$v\ll1$, $E_R\ll M_A$, and $\delta\ll m$, the minimum incident
speed for a nuclear recoil of energy $E_R$ is
\begin{equation}
v_{\min,A}(E_R)
=
\frac{M_AE_R/\mu_A+\delta}{\sqrt{2M_AE_R}},
\label{eq:minimum_speed}
\end{equation}
where $\mu_A=mM_A/(m+M_A)$ is the ALP--nucleus reduced mass.

Minimizing $v_{\min,A}(E_R)$ with respect to $E_R$ gives the
threshold speed
\begin{equation}
v_{\mathrm{thr},A}=\sqrt{\frac{2\delta}{\mu_A}}.
\label{eq:excitation_threshold}
\end{equation}

For $v>v_{\mathrm{thr},A}$, the allowed recoil energies
satisfy $v_{\min,A}(E_R)\leq v$. Solving
$v_{\min,A}(E_R)=v$ for $E_R$ gives the lower and upper
endpoints of this interval,
\begin{equation}
E_{R,A}^{\pm}(v)
=
\frac{\mu_A^2v^2}{2M_A}
\left(
1\pm\sqrt{1-\frac{2\delta}{\mu_Av^2}}
\right)^2.
\label{eq:recoil_endpoints}
\end{equation}
At threshold, the two endpoints coincide at
$E_{R,A}^{*}=\mu_A\delta/M_A$. For a halo distribution with
maximum laboratory speed $v_{\max}$, the kinematics requires $\delta<\mu_Av_{\max}^2/2$. This condition is applied separately to each isotope because $\mu_A$ depends on the nuclear mass.

The leading non-relativistic cross section (The derivation is given in Appendix~\ref{app:amplitude}.) for coherent scattering from the nuclear is
\begin{align}
\frac{\mathrm{d}\sigma_A}{\mathrm{d}E_R}
={}&
\frac{\alpha G_{a\gamma\gamma^{\prime}}^{\,2}Z^2}
{4E_R}\,F_A^2(Q)
\nonumber\\
&\times
\left(1-\frac{v_{\min,A}^2(E_R)}{v^2}\right)
\Theta\!\left(v-v_{\min,A}(E_R)\right),
\label{eq:differential_cross_section}
\end{align}
where $\alpha=e^2/(4\pi)$ is the fine-structure constant and $\Theta$ is the Heaviside step function.The factor
$1-v_{\min,A}^2(E_R)/v^2$ arises from the transverse-momentum structure of the antisymmetric portal vertex and vanishes at both recoil endpoints. $F_A(Q)$ is the nuclear charge form factor with $Q\simeq\sqrt{2M_AE_R}$.
For the numerical analysis, we use the Helm parametrization~\cite{Helm:1956zz}
\begin{equation}
F_A(Q)=\frac{3j_1(QR_A)}{QR_A}
\exp\!\left[-\frac{(Qs)^2}{2}\right],
\label{eq:form_factor}
\end{equation}
where $j_1$ is the spherical Bessel function of order one,
$R_A=1.14\,A^{1/3}\,\mathrm{fm}$ is the radius parameter,
$A$ is the nuclear mass number, and $s=0.9\,\mathrm{fm}$
is the surface-smearing length. The form factor satisfies
$F_A(0)=1$.
\subsection{Boosted DM Scattering}

In this subsection, we derive the kinematics of boosted DM transition. The incoming ALP and the outgoing dark photon have
energies $E_1=p^0$ and $E_2=p^{\prime 0}$, respectively. The energy of the final nucleon is
\begin{equation}
E_{A'} = \frac{1}{2m_{A}}\big(2p\cdot p'+2m^2_{A}-m^2_{1}-m^2_{2}\big).
\end{equation}
Because the nucleon is slow enough, the recoiling kinetic energy is $E_{R}\simeq E_{A'}-m_{A}$. Therefore,
\begin{equation}\label{eq:E_nr}
E_{R}= \frac{1}{2m_{A}}\big(2p\cdot p'-m^2_{1}-m^2_{2}\big).
\end{equation}
Denoting $E^{*}_{1,2},~p^*_{1,2}$ as the energy and momentum magnitudes under CM frame before and after scattering respectively, the recoiling energy is
\begin{equation}
E_{R}= \frac{1}{2m_{A}}\big(2E^*_{1} E^*_{2}-2p^*_{1}p^*_{2}\cos\theta_{1,2}-m^2_{1}-m^2_{2}\big).
\end{equation}
The energy bounds are thus
\begin{equation}
E^{\pm}_{R,\rm exac}(s) = \frac{1}{2m_{A}}\big(2E^*_{1} E^*_{2}\pm 2p^*_{1}p^*_{2}-m^2_{1}-m^2_{2}\big),
\end{equation}
where
\begin{equation}
E^*_{1,2} = \frac{s+m^2_{1,2}-m^2_{A}}{2\sqrt{s}},\quad p^*_{1,2}=\frac{\sqrt{\lambda(s,m^2_{1,2},m^2_{A})}}{2\sqrt{s}}.
\end{equation}
The symbol $\pm$ stands for upper and lower bound respectively. The Lorentz invariant form of the two bodies cross section is
\begin{equation}
\begin{aligned}
d\sigma^{C}_{A} =& \frac{1}{4\sqrt{(k\cdot p)^2-m^2_{1}m^2_{A}}}(2\pi)^4\delta^{4}(p+k-p'-k')
\\
&\times\overline{|\mathcal{M}|}^2
\frac{d^3{\bf p'}}{(2\pi)^3 2E_{2}}
\frac{d^3{\bf k'}}{(2\pi)^3 2E_{A'}},
\end{aligned}
\end{equation}
the explicit form of $\overline{|\mathcal{M}|}^2$  is shown in Appendix~\ref{app:amplitude}. The Møller flux factor can be written as
\begin{equation}
4\sqrt{(k\cdot p)^2-m^2_{1}m^2_{A}}=2\sqrt{\lambda(s,m^2_{1},m^2_{A})}.
\end{equation}

According to the endothermic process, there are
\begin{equation}\label{eq:threshold_s}
s=(p+k)^2\geq(m_2+m_{A})^2.
\end{equation}
The scattering can happen when the initial energy satisfying
\begin{equation}
\label{eq:E_thres}
E_{1}\geq E_{1,\rm th}=m_{2}+\frac{m_2^2-m^2_1}{2M_{A}},
\end{equation}
this threshold varies among different isotopes. The recoiling energy reads
\begin{equation}
E_{R} = \frac{m^2_{2}-m^2_{1}}{2(M_{A}+m_{2})},
\end{equation}
when the equal symbol is taken and, when $E_{1}>E_{1,\rm th}$, there are $E^{-}_{R,\rm exac}\leq E_{R}\leq E^{+}_{R,\rm exac}$. Therefore, the cross section can be written as a function of recoiling energy
\begin{equation}\label{eq:invariant_cross_section}
d\sigma^{C}_{A} = \frac{M_A}{8\pi \lambda(s,m_{1}^2,M^2_{A})}\overline{|\mathcal{M}|}^2 \Theta_A^{\mathrm{exact}}dE_{R},
\end{equation}
where, $E_{R}$ is given by Eq.~\eqref{eq:E_nr} and squared amplitude is given by Eq.~\eqref{eq:sqM_axion}. The Heaviside function $\Theta_A^{\mathrm{exact}}$ means when $E_1\geq E_{1,\mathrm{th}}^A$ and
$E_{R,\mathrm{exact}}^-\leq E_R \leq E_{R,\mathrm{exact}}^+$, it equals unit, and vanishes otherwise. Its leading non-relativistic limit is $\Theta_A$ used in Eq.~\eqref{eq:differential_cross_section}.
\section{Event rates and LZ statistical analysis}
\label{sec:statistic}
\subsection{Event rates of non-relativistic DM}
\label{NR_analyze}
We assume that the incident ALPs follow the SHM, with a local density
$\rho_a=0.3\,\mathrm{GeV\,cm^{-3}}$.
The laboratory-frame speed distribution $f_{\mathrm{lab}}(v)$
is obtained from a truncated Galactic Maxwell distribution
\cite{Freese:2012xd} and satisfies
$\int_0^{v_{\max}}f_{\mathrm{lab}}(v)\,\mathrm{d}v=1$.
We adopt a Galactic velocity scale
$v_0=220\,\mathrm{km\,s^{-1}}$, an escape speed
$v_{\mathrm{esc}}\simeq542.62\,\mathrm{km\,s^{-1}}$,
and a fixed laboratory speed
$v_E\simeq218.85\,\mathrm{km\,s^{-1}}$,
giving $v_{\max}=v_{\mathrm{esc}}+v_E
\simeq761.47\,\mathrm{km\,s^{-1}}$.

The recoil rate is obtained by weighting the differential
cross section in Eq.~\eqref{eq:differential_cross_section}
with the incident ALP number density $\rho_a/m$ and speed
$v$, and averaging over the speed distribution.
Summing over the xenon isotopes gives the differential rate
per unit detector mass and unit time,
\begin{equation}
\frac{\mathrm{d}R}{\mathrm{d}E_R}
=
\frac{\rho_a}{m}
\frac{N_{\mathrm{Av}}}{\overline{M}_{\mathrm{mol}}}
\sum_A x_A
\int_0^{v_{\max}}\mathrm{d}v\,
f_{\mathrm{lab}}(v)\,v\,
\frac{\mathrm{d}\sigma_A}{\mathrm{d}E_R},
\label{eq:differential_recoil_rate}
\end{equation}
where $N_{\mathrm{Av}}$ is Avogadro's constant,
$\overline{M}_{\mathrm{mol}}$ is the mean xenon molar mass,
and $x_A$ is the natural number fraction of isotope $A$,
with $\sum_A x_A=1$.
The factor $x_A N_{\mathrm{Av}}/\overline{M}_{\mathrm{mol}}$
therefore counts the target nuclei of isotope $A$ per unit
detector mass. Each isotope contributes with its own
nuclear mass, charge form factor, and kinematic threshold.

Multiplying the rate by the exposure $\mathcal{E}$ gives the
expected true recoil spectrum,
\begin{equation}
\frac{\mathrm{d}N_{\mathrm{true}}}{\mathrm{d}E_R}
=
\mathcal{E}\,
\frac{\mathrm{d}R}{\mathrm{d}E_R},
\label{eq:true_recoil_spectrum}
\end{equation}
where we use $\mathcal{E}=2.84\,\mathrm{tonne\,yr}$.
This spectrum describes the expected nuclear recoils before
detector efficiency and energy resolution are applied.

To obtain the observable spectrum, we account for the
detection efficiency $\epsilon(E_R)$ and the finite
resolution of the reconstructed recoil energy $E_d$.
Following Ref.~\cite{Fan:2026hzw}, the probability density for
reconstructing an energy $E_d$ from a true recoil $E_R$
is modeled by
\begin{equation}
K(E_d,E_R)
=
\frac{1}{\sqrt{2\pi}\sigma_E(E_R)}
\exp\!\left[
-\frac{(E_d-E_R)^2}{2\sigma_E^2(E_R)}
\right],
\label{eq:response_kernel}
\end{equation}
where $\sigma_E(E_R)$ is the energy resolution,
\begin{equation}
\frac{\sigma_E(E_R)}{\mathrm{keV}}
=
\sqrt{
1.46^2\frac{E_R}{\mathrm{keV}}
+
\left(0.093\frac{E_R}{\mathrm{keV}}\right)^2
}.
\label{eq:energy_resolution}
\end{equation}
The reconstructed spectrum is therefore
\begin{equation}
\frac{\mathrm{d}N}{\mathrm{d}E_d}
=
\int_0^\infty\mathrm{d}E_R\,
\epsilon(E_R)K(E_d,E_R)
\frac{\mathrm{d}N_{\mathrm{true}}}{\mathrm{d}E_R}.
\label{eq:reconstructed_spectrum}
\end{equation}

\subsection{Event rates of boosted DM}

Generally, the explicit boosted DM differential rate per detector mass takes the form~\cite{Zhang:2023oed}
\begin{equation}
\frac{dR_A}{dE_{R}} = \int^{E^{max}_{S}}_{E^{min}_{S}}\frac{d\Phi}{dE_{S}}\frac{d\sigma_{A}}{dE_{R}}dE_{S},
\end{equation}
here, we parametrize the initial flux in following way
\begin{equation}
\frac{d\Phi}{dE_{S}}=\Phi_{0}f(E_{S};E_{c},\sigma_{\rm in}).
\end{equation}
The incoming DM is modeled as a Gaussian distribution on incident energy~\cite{Acevedo:2026tur}.
The distribution function $f$ is
\begin{equation}
f(E_{S}; E_{c},\sigma_{\rm in})=N_{E_{c},\sigma_{\rm in}} \exp\bigg[{\frac{(E_{S}-E_{c})^2}{2\sigma^2_{\rm in}}}\bigg],
\end{equation}
as the width $\sigma_{\rm in}$ tends to zero, it comes to monochromatic case~\cite{Agashe:2014yua}
\begin{equation}
f(E_{S};E_{c},\sigma_{\rm in})=\delta(E_{S}-E_{c}).
\end{equation}
Introducing an energy resolution, the reconstructed differential event distribution is
\begin{equation}\label{eq:boosted_event}
\begin{aligned}
\frac{dN}{dE_{d}}=& \mathcal{E}\frac{N_{{\rm Av}}}{\overline{M}_{\rm mol}}\sum_{A} x_{A}\int_{0}^{\infty}dE_{S}~f(E_S; E_{c},\sigma_{\rm in})
\\
&\times\int_{0}^{\infty}dE_{R}~K(E_{d};E_{R})\epsilon(E_{R})\frac{d\sigma_{A}}{dE_{R}}.
\end{aligned}
\end{equation}
Other functions as well as parameters here have been defined previously.

\subsection{Statistical methods}\label{Likelihood}
According to the detecting efficiency in Ref.\cite{LZ:2026axp}, we have taken the event-region as
\begin{equation}
E_{R}\in [215.5~\keV, 280.5~\keV],
\end{equation}
while the total $E_{R}$ for analysis is
\begin{equation}
E_{R}\in [70~\keV, 305~\keV].
\end{equation}
In following numerical analysis, we have reconstructed the energy bins as
\begin{align}
[E_1^-,E_1^+]&=[70,215.5]\,\mathrm{keV},
\nonumber\\
[E_2^-,E_2^+]&=[215.5,280.5]\,\mathrm{keV},
\nonumber\\
[E_3^-,E_3^+]&=[280.5,305]\,\mathrm{keV},
\label{eq:bin_edges}
\end{align}
where $E_i^-$ and $E_i^+$ are the lower and upper boundaries
of bin $i$. The expected number of events in each bin is
\begin{equation}
N_i
=
\int_{E_i^-}^{E_i^+}\mathrm{d}E_d\,
\frac{\mathrm{d}N}{\mathrm{d}E_d}.
\label{eq:bin_event_counts}
\end{equation}
Energy smearing allows accepted recoils with $E_R<300\,
\mathrm{keV}$ to contribute to reconstructed energies above 300 keV.

Here we have adopted a reduced Poisson likelihood function $\mathcal{L}$ neglecting the background details.
\begin{equation}\label{eq:Poisson_Like}
\mathcal{L}=\prod_{i}\frac{N_{i}^{k_{i}}}{k_{i}!}e^{-N_{i}},
\end{equation}
where $k_{i}$ stands for the observed event numbers and $N_{i}$ the expected event numbers at $i$-th bin. Here we have used three-bin analysis \cite{Fan:2026hzw} with assigned counts $(k_1,k_2,k_3)=(0,1,0)$. Therefore the likelihood function becomes
\begin{equation}
\mathcal{L}=N_{2}e^{-N_{\rm tot}},
\end{equation}
with $N_{\rm tot}=\sum_{i}N_{i}$. This likelihood incorporates both the event in the middle bin and the absence of events in the outer bins.

The Poisson deviance function is defined as:
\begin{equation}
 \chi_P^2=
 -2\ln\frac{\mathcal L}{\mathcal L_{\rm sat}}
 =
 2\left(N_{\rm tot}-1-\ln N_2\right),
 \label{eq:poisson_deviance}
\end{equation}
where $\mathcal L_{\rm sat}=e^{-1}$ is the saturated-model likelihood obtained at $N_i=k_i$. The minimum of this function means the best-fit point in parameter space. In case of non-relativistic analysis, we compute the expected counts using the laboratory-frame SHM distribution. The integral accounts for the speed-dependent flux and the scattering threshold of each xenon isotope. The signal counts scale as $N_i\propto G_{a\gamma\gamma'}^2$. Varying this coupling therefore changes only the signal normalization. In the case of $N_{\rm tot}\simeq 1$, there is: $N_{\rm tot}-1\simeq {\rm ln}~N_{\rm tot}$, then Eq.~\eqref{eq:poisson_deviance} can be rewritten as
\begin{equation}
 \chi_{P,\mathrm{prof}}^2=-2\ln f_2,
 \qquad
 f_2=\frac{N_2}{N_{\rm tot}}.
 \label{eq:shape_deviance}
\end{equation}
In this case, the fit quality depends only on the middle-bin fraction. 
For boosted DM analysis, the corresponding expected differential event numbers are calculated by Eq.~\eqref{eq:boosted_event} and the expected count $N_i$ in each bin scales as $\Phi_0 G_{a\gamma\gamma'}^2$.

For each fixed $(m,\delta)$, we minimize $\chi^2_{P}$ over the allowed values of the remaining parameter set $\mathcal{P}$ and define
\begin{equation}
 \Delta\chi^2(m,\delta)=
 \min_{\mathcal{P}}
 \chi_P^2(m,\delta,\mathcal{P})
 -
 \chi_{P,\min}^2,
 \label{eq:delta_chi_squared}
\end{equation}
where $\chi_{P,\min}^2$ is the global minimum over the adopted parameter ranges. Here, $\mathcal{P}=\{G_{a\gamma\gamma'}\}$ for the non-relativistic case and $\mathcal{P}=\{E_c,\sigma_{\rm in}\}$ for the boosted case. We use $\Delta\chi^2=2.28$ and $4.61$ to define nominal 68\% and 90\% confidence contours in the two-dimensional $(m,\delta)$ plane. These asymptotic contours should be interpreted as approximate for a sample containing only a single event.

\FloatBarrier

\section{Results}\label{Sec:result}
\subsection{Parameter space and benchmark}
For the non-relativistic analysis, we scan over the following parameter ranges
\begin{equation}
\begin{gathered}
400\leq m/\mathrm{GeV}\leq1200,
\qquad
100\leq\delta/\mathrm{keV}\leq1000,
\\
10^{-8}
\leq
G_{a\gamma\gamma^{\prime}}/
(\mathrm{GeV}^{-1})
\leq10^{-1}.
\end{gathered}
\label{eq:s4_domain}
\end{equation}
The local DM density, exposure, and nuclear-recoil response are fixed as specified in Sec.~\ref{NR_analyze}.
The incident DM speed is integrated over the normalized SHM distribution.
For each pair $(m,\delta)$, we choose the value of $G_{a\gamma\gamma^{\prime}}$ so that one-event can be approximately produced while minimizing $\chi_P^2$
within the scan range.

TABLE~\ref{tab:benchmark} summarizes the best-fit benchmark. It gives approximately one expected event in the three reconstructed-energy bins, with $60.75\%$ of the signal in the middle bin.
The minimum Poisson deviance is $\chi_{P,\min}^2=0.99835$. Both the ALP mass and the portal coupling reach
their upper scan limits.
\begin{table}[htbp]
\centering
\caption{
Best-fit benchmark for the fixed SHM and
nuclear-recoil response adopted in Sec.~\ref{NR_analyze}.
The counts include an exposure of
$2.84\,\mathrm{tonne\,yr}$.
Here $f_2$ has been defined in Eq.~\ref{eq:shape_deviance}.
}
\label{tab:benchmark}
\begin{ruledtabular}
\begin{tabular}{lc}
Quantity & Value \\
\hline
$m$ & $1200\,\mathrm{GeV}$ \\
$\delta$ & $346.119\,\mathrm{keV}$ \\
$G_{a\gamma\gamma^{\prime}}$
& $0.1000\,\mathrm{GeV}^{-1}$ \\
$N_1$ & $0.31354$ \\
$N_2$ & $0.58466$ \\
$N_3$ & $0.06425$ \\
$N_{\rm tot}$ & $0.96245$ \\
$f_2$ & $60.75\%$ \\
$\chi_P^2$ & $0.99835$ \\
\end{tabular}
\end{ruledtabular}
\end{table}
We weight xenon isotopes by their natural abundances. At the benchmark, scattering on ${}^{124}\mathrm{Xe}$ and
${}^{126}\mathrm{Xe}$ is kinematically forbidden. The signal comes from the high-speed tail of the SHM and is dominated by ${}^{132}\mathrm{Xe}$, ${}^{134}\mathrm{Xe}$, and ${}^{136}\mathrm{Xe}$.

\subsection{Recoil spectra and three-bin event counts}

\begin{figure}[htbp]
\centering
\includegraphics[width=\columnwidth]{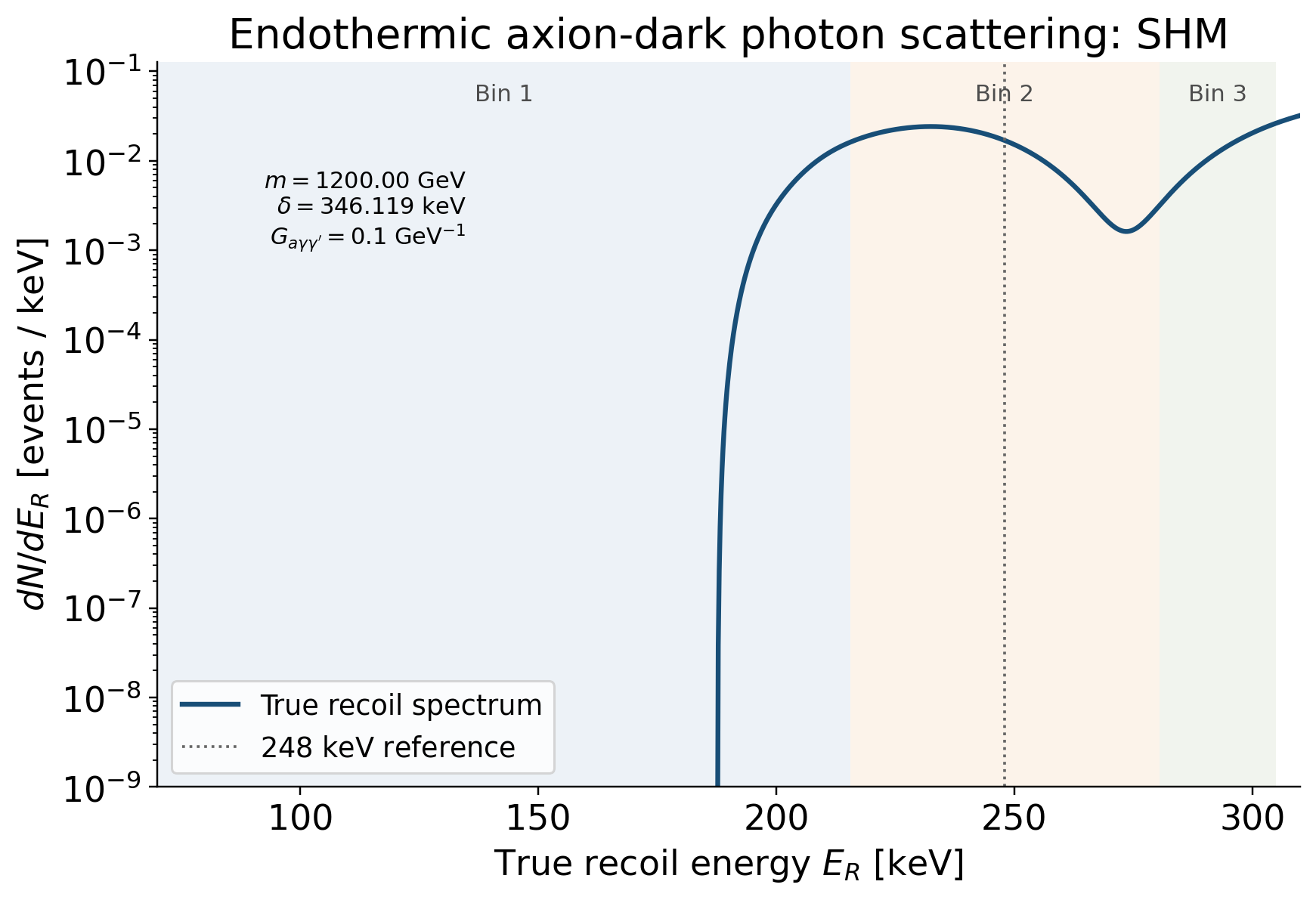}
\caption{
True recoil spectrum at the benchmark in
TABLE~\ref{tab:benchmark}, including the exposure.
The shaded regions indicate the numerical energy
boundaries of the three LZ analysis bins for
reference on the $E_R$ axis.
The dotted line marks $248\,\mathrm{keV}$.
}
\label{fig:true}
\end{figure}

FIG.~\ref{fig:true} shows the true recoil spectrum
$\mathrm{d}N_{\mathrm{true}}/\mathrm{d}E_R$ at the benchmark.
We compute the recoil rate for each xenon isotope using the
SHM distribution, sum the rates weighted by their natural
abundances, and multiply the total by the exposure.
The spectrum has two maxima near $E_R=232$ and $339\,\mathrm{keV}$.
Its shape follows from the velocity-averaged scattering cross
section and the nuclear charge form factors.

\begin{figure}[htbp]
\centering
\includegraphics[width=\columnwidth]
{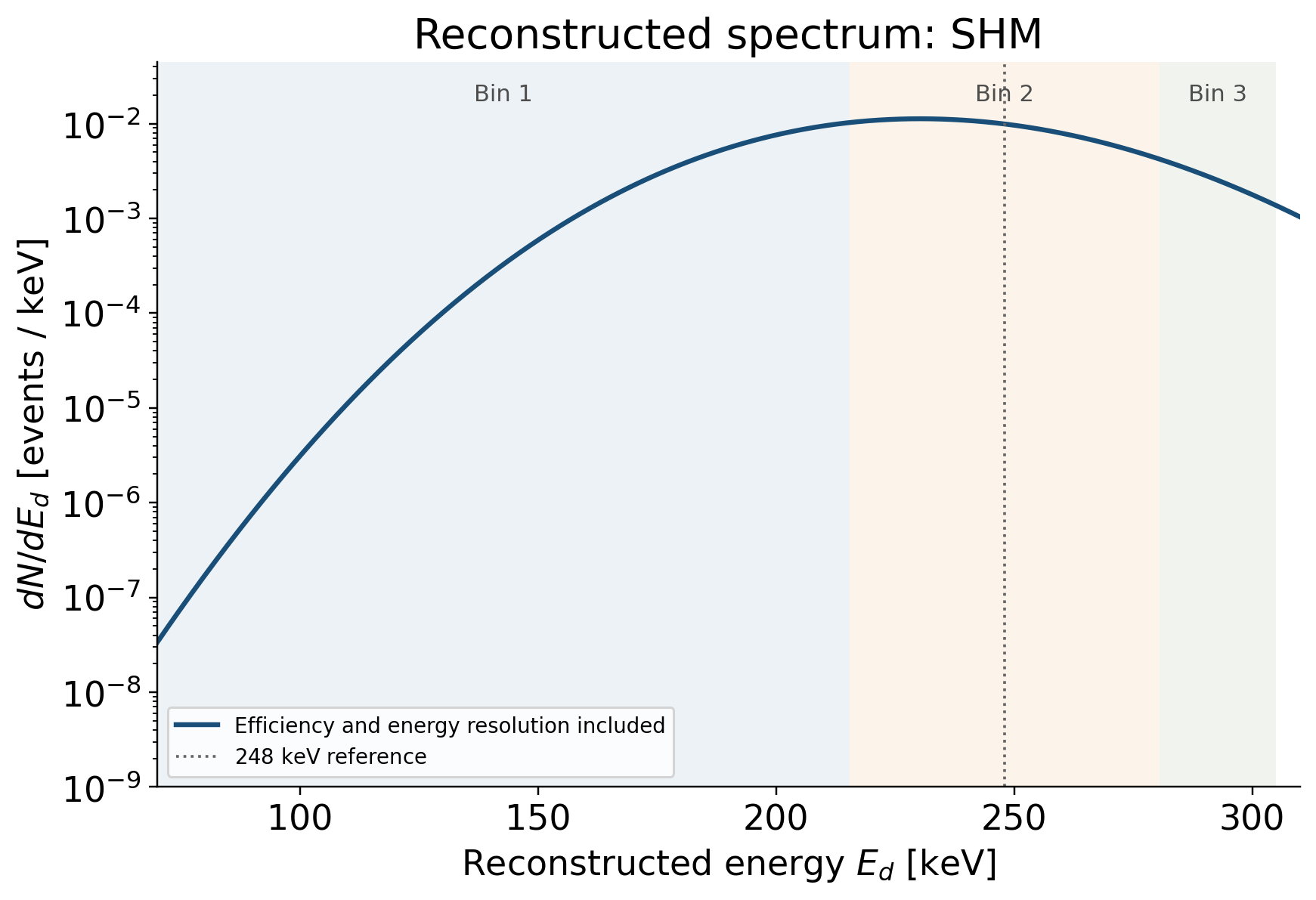}
\caption{
Reconstructed spectrum
$\mathrm{d}N/\mathrm{d}E_d$ at the same benchmark,
including detection efficiency and energy
resolution.
}
\label{fig:reconstructed}
\end{figure}

FIG.~\ref{fig:reconstructed} shows the reconstructed recoil
spectrum $\mathrm{d}N/\mathrm{d}E_d$ at the benchmark.
We multiply the true recoil spectrum by the detection efficiency
and convolve the result with the Gaussian energy response.
The reconstructed spectrum peaks near $E_d=230\,\mathrm{keV}$
in the middle bin.

\begin{figure}[htbp]
\centering
\includegraphics[width=\columnwidth]
{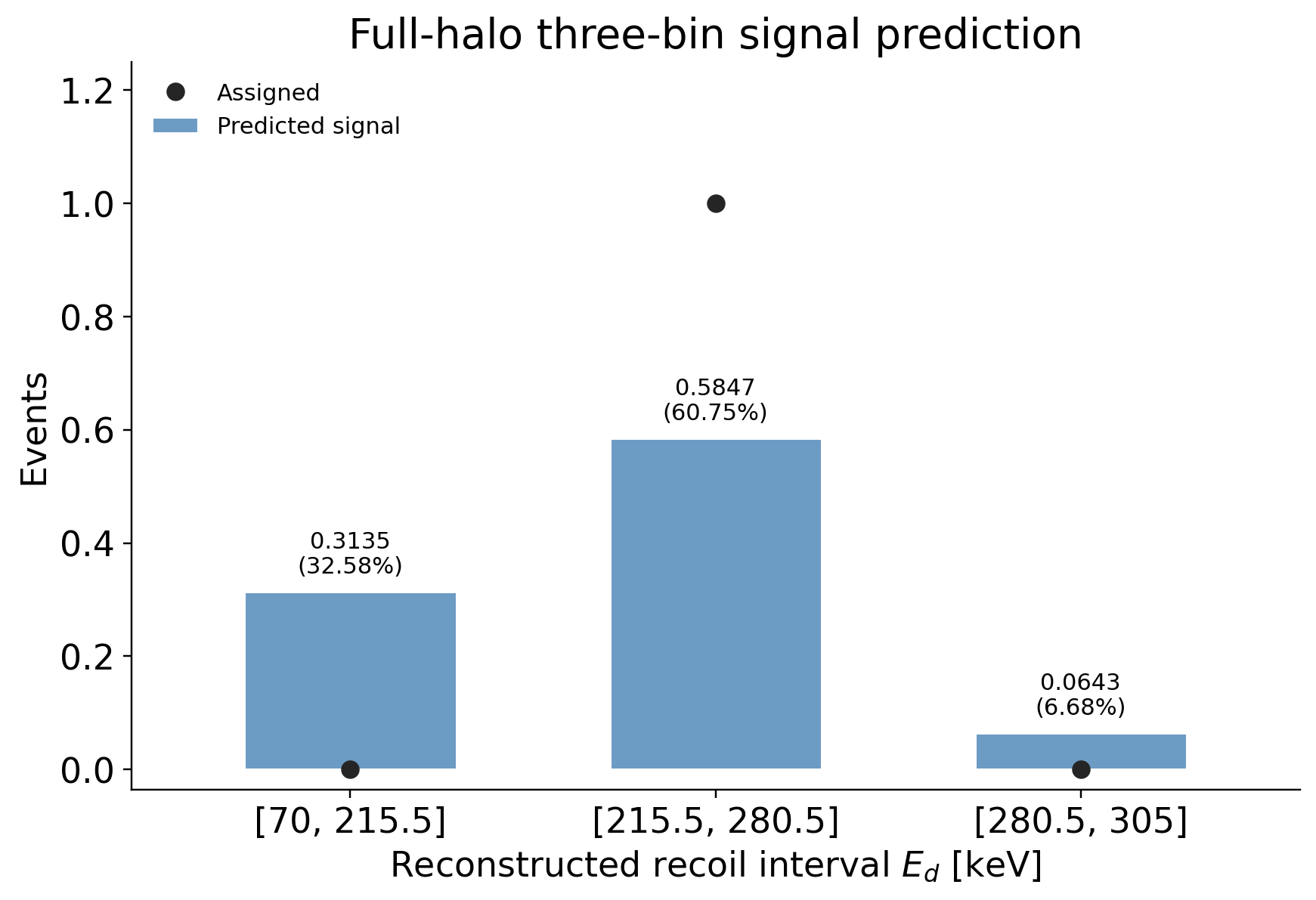}
\caption{
Expected event counts in the three reconstructed
energy bins at the benchmark.
The bars show the predicted means, and the black
markers show the assigned counts
$(0,1,0)$.
The labels above the bars give the expected
counts and their fractions of the three-bin
signal.
}
\label{fig:bins}
\end{figure}

FIG.~\ref{fig:bins} compares the expected event counts at the benchmark with the assigned counts $(k_1,k_2,k_3)=(0,1,0)$ used in the Poisson fit. We calculate the expected count $N_i$ in each bin using Eq.~\eqref{eq:bin_event_counts}. The middle bin dominates the predicted signal, with $N_2=0.58466$ expected events, accounting for $60.75\%$ of the total.

\subsection{Profile likelihood in the mass-splitting plane}

\begin{figure}[htbp]
\centering
\includegraphics[width=\columnwidth]
{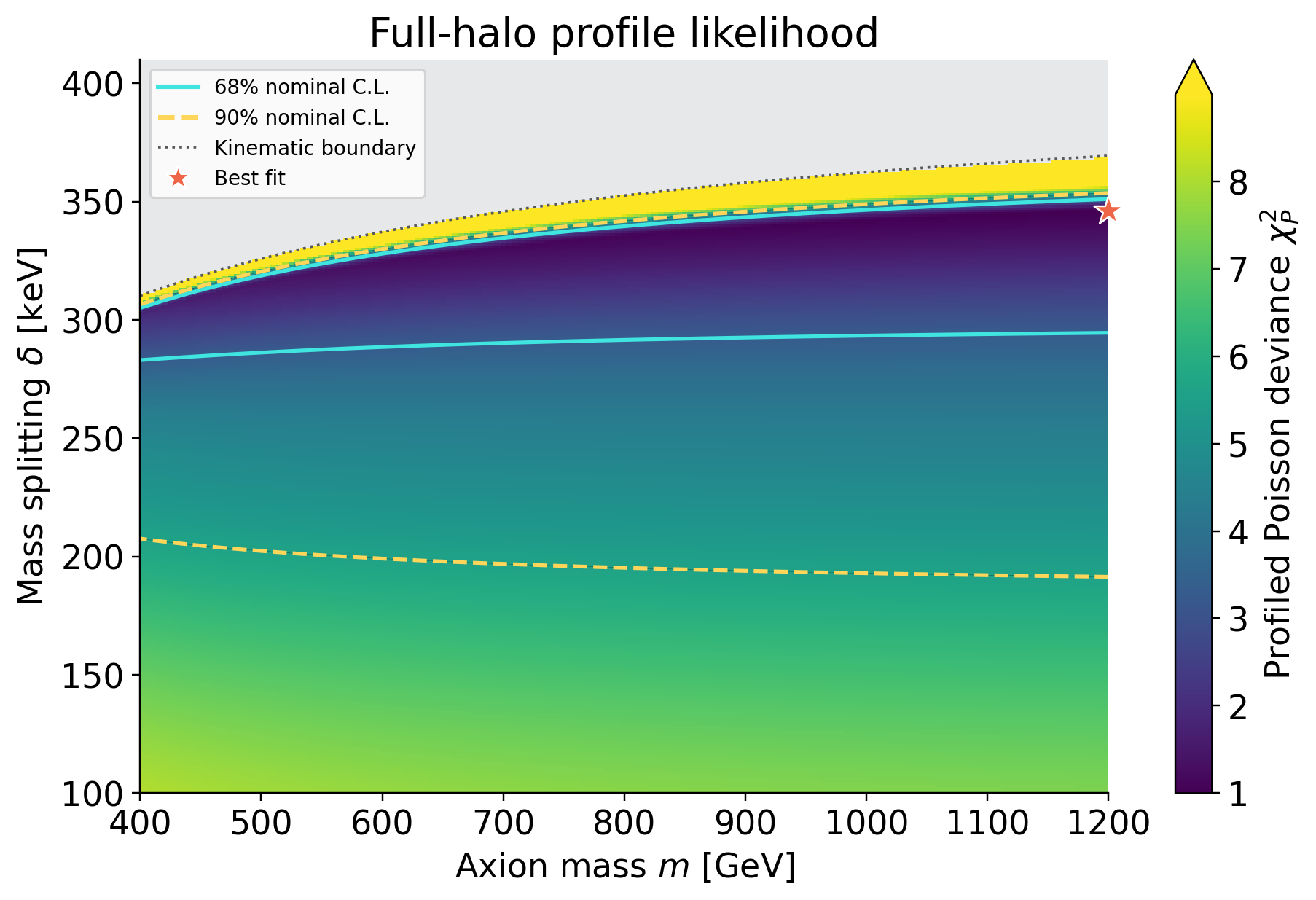}
\caption{Profile likelihood in the ALP mass-splitting plane. The color scale shows the absolute Poisson
deviance after profiling the portal coupling. The solid and dashed contours indicate the
nominal $68\%$ and $90\%$ two-parameter regions. The star marks the benchmark in TABLE~\ref{tab:benchmark}.
The dotted curve is the ${}^{136}\mathrm{Xe}$ kinematic boundary, with the kinematically forbidden region shaded
in gray.}
\label{fig:profile}
\end{figure}

FIG.~\ref{fig:profile} displays the profile likelihood in the $(m,\delta)$ plane after optimizing $G_{a\gamma\gamma^{\prime}}$. The color scale represents the absolute profiled Poisson deviance $\chi_P^2$.
The nominal $68\%$ and $90\%$ contours correspond to $\Delta\chi_P^2=2.2789$ and $4.6052$, respectively, relative to the benchmark minimum.
A good fit favors larger mass splittings as the ALP mass increases. A heavier ALP gives a larger ALP--nucleus reduced mass, so the same incident speed allows a larger splitting. 
The heaviest xenon isotope ${}^{136}\mathrm{Xe}$ sets the upper splitting limit.
Above this boundary, all xenon scattering channels are closed.

At fixed $m$, on the one hand, small splittings allow lower-energy recoils
and shift more of the signal into the first bin. On the other
hand, a splitting near the kinematic limit restricts scattering
to the fastest halo DMs and yields too few events even
at the largest coupling. 
The best-fit point has an intermediate mass splitting of $\delta\simeq346\,\mathrm{keV}$, which suppresses low-energy recoils. The total expected count is close to one, with the middle bin dominating.
At fixed recoil energy $E_R$, the differential cross section
vanishes below $v_{\min,A}(E_R)$ and increases with speed as
$1-v_{\min,A}^2(E_R)/v^2$ above it. Scattering just above the
threshold is therefore suppressed, and the xenon charge form
factor further reduces the rate. Since the event count scales
as $G_{a\gamma\gamma^{\prime}}^2$, a coupling of
$0.1\,\mathrm{GeV}^{-1}$ brings the expected total to about one
event at the benchmark. Experimental constraints on the coupling are relatively weak in this high-mass regime~\cite{Chen:2025hbh,Yue:2026oqu,Chen:2026beb,Guha:2018mli}, while the theoretical status of this benchmark remains an open question.

\subsection{Boosted DM analysis}
In this section, the analysis on boosted DM transition is shown. We focus on incident speeds above the halo cutoff $v_{0}>v_{\rm max}$. The sampled speeds are $v_{0}/c=0.005,0.01,0.05$ and $0.1$ approximately two to forty times $v_{\max}$. 
We consider a monoenergetic spectrum and a Gaussian spectrum
with mean $E_c$ and width $\sigma_{\rm in}$ for the incident DM. In the Gaussian case, $v_0$ denotes the speed associated with $E_c$. We adopt $G_{a\gamma\gamma'}\sim10^{-4}\,\mathrm{GeV}^{-1}$ in this
subsection, guided by laboratory searches for ALPs and dark photons
\cite{Chen:2025hbh,Yue:2026oqu,Chen:2026beb} and astrophysical
constraints \cite{Guha:2018mli}.

\begin{figure}[t]
    \centering
    \includegraphics[width=0.9\linewidth]{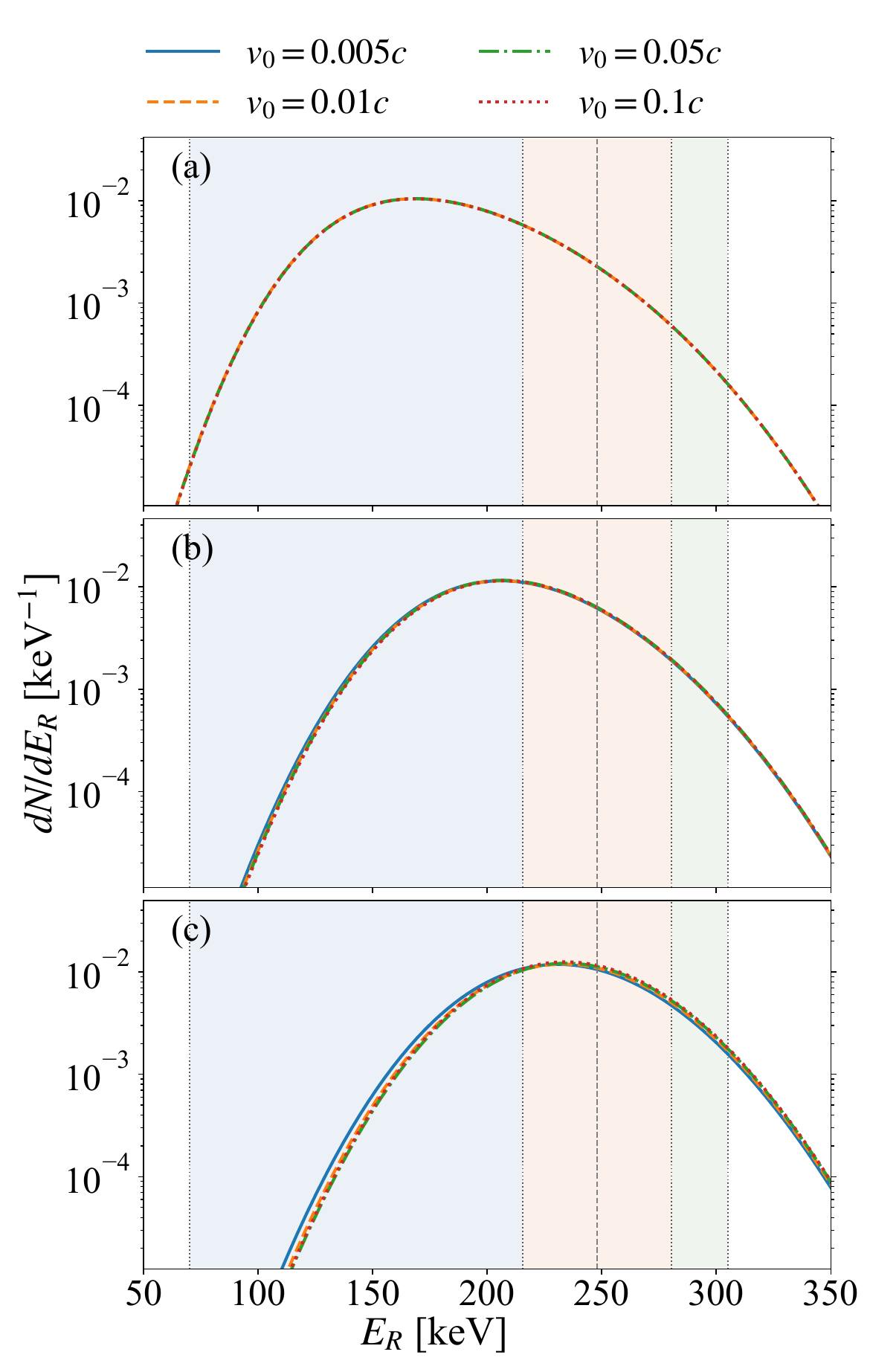}
    \caption{Recoil-energy spectra for monoenergetic boosted DM. Panels (a), (b), and (c) show fluxes $\Phi_0=10^3$, $10^4$, and $10^5\,\mathrm{cm}^{-2}\mathrm{s}^{-1}$, respectively. The blue solid, orange dashed, green dot-dashed, and red dotted curves correspond to $v_0/c=0.005$, $0.01$, $0.05$, and $0.1$, respectively. The benchmark parameters are listed in Table~\ref{tab:bf_fix_lzE}.}
    \label{fig:fixedv_spectrum}
\end{figure}
\begin{table}[t]
\centering
\caption{Best-fit parameters and expected event numbers for monochromatic
initial velocity. Mass of initial state $m$ are given in GeV and mass splitting $\delta$ in MeV. The initial flux $\Phi_{0}$ is in unit of ${\rm cm}^{-2}{\rm s}^{-1}$. The red value of $\chi^2_{P}$ is the minimum among the TABLE.}
\label{tab:bf_fix_lzE}
\vspace{4pt}
\setlength{\tabcolsep}{3pt}
\begin{tabular}{c c ccccc}
\toprule[0.6pt]        
$\Phi_0$ & $v_0$ & $m$ & $\delta$ & $\chi^2_{P}$ & $N_{\rm tot}$ & $N_{2}$ \\
\midrule[0.3pt]       
\multirow{4}{*}{$10^{5}$} & 0.005 & 99.95 & 0.6668 & 0.977 & 1.05 & 0.624 \\
 & 0.01 & 35.43 & 1.332 & 0.909 & 1.04 & 0.638 \\
 & 0.05 & 5.739 & 6.659 & {\color{red}0.877} & 1.05 & 0.655 \\
 & 0.1 & 2.793 & 13.34 & 0.880 & 1.08 & 0.674 \\
\midrule[0.3pt]       
\multirow{4}{*}{$10^{4}$} & 0.005 & 103.9 & 0.653 & 1.805 & 1.02 & 0.412 \\
 & 0.01 & 36.38 & 1.305 & 1.778 & 1.02 & 0.415 \\
 & 0.05 & 5.864 & 6.524 & 1.768 & 1.02 & 0.42 \\
 & 0.1 & 2.852 & 13.08 & 1.759 & 1.01 & 0.417 \\
\midrule[0.3pt]       
\multirow{4}{*}{$10^{3}$} & 0.005 & \multicolumn{5}{c}{\ding{55}} \\
 & 0.01 & 39.49 & 1.226 & 3.573 & 1.01 & 0.169 \\
 & 0.05 & 6.26 & 6.131 & 3.570 & 1.01 & 0.169 \\
 & 0.1 & 3.039 & 12.29 & 3.568 & 1 & 0.169 \\
\midrule[0.3pt]       
\multirow{4}{*}{$10^{2}$} & 0.005 & \multicolumn{5}{c}{\ding{55}} \\
 & 0.01 & \multicolumn{5}{c}{\ding{55}} \\
 & 0.05 & \multicolumn{5}{c}{\ding{55}} \\
 & 0.1 & \multicolumn{5}{c}{\ding{55}} \\
\bottomrule[0.6pt]     
\end{tabular}
\end{table}

FIG.~\ref{fig:fixedv_spectrum} shows the recoil-energy spectra of a mono-energetic initial state, which differs only in the assumed flux. The shaded bands with different color label the energy bins defined in Eq.~\eqref{eq:bin_edges}. 
TABLE~\ref{tab:bf_fix_lzE} lists the best-fit mass parameters for each spectrum at the specified $v_0$ and $\Phi_0$. At fixed total event count, the Poisson deviance $\chi_P^2$ describes the spectral shape, including the location of the peak, through the relative counts in the three bins. The coupling and flux must be constrained independently because the overall rate depends on their combination $\Phi_0 G_{a\gamma\gamma'}^2$. The best-fit spectra give approximately one expected event in total, except for the $\Phi_0=10^{2}\,\mathrm{cm}^{-2}\,\mathrm{s}^{-1}$ case.

FIG.~\ref{fig:fixedv_spectrum} and TABLE~\ref{tab:bf_fix_lzE} show that, as $\Phi_0$ increases, the best-fit spectra peak closer to or within the signal window. At low flux, none of the speeds gives enough events. 
A lower speed generally leads to further suppression of the factor $|{\bf p}\times {\bf q}|^2$ in Eq.~\eqref{eq:app_lab_squared_amplitude}. Therefore, as $\Phi_{0}$ decreases, the small-$v_0$ channel closes first. Spectra with larger deviances tend to peak at lower recoil energies. This is because, at each $v_0$, the fitted $m$ and $\delta$ must
satisfy the incident-energy threshold in
Eq.~\eqref{eq:E_thres}. Near this threshold, a smaller mass gives a lower recoil energy at fixed $\delta$. When the incident DM is boosted, the mass can be lower. This brings the energy thresholds for different isotopes closer together, allowing the benchmarks in TABLE.~\ref{tab:bf_fix_lzE} to open scattering channels of all isotopes. However, because the incident energy remains close to the thresholds, Eq.~\eqref{eq:app_lab_squared_amplitude} is still significantly suppressed, the cross section is still small at these benchmark points.

\begin{figure}[t]
    \centering
    \includegraphics[width=\columnwidth]{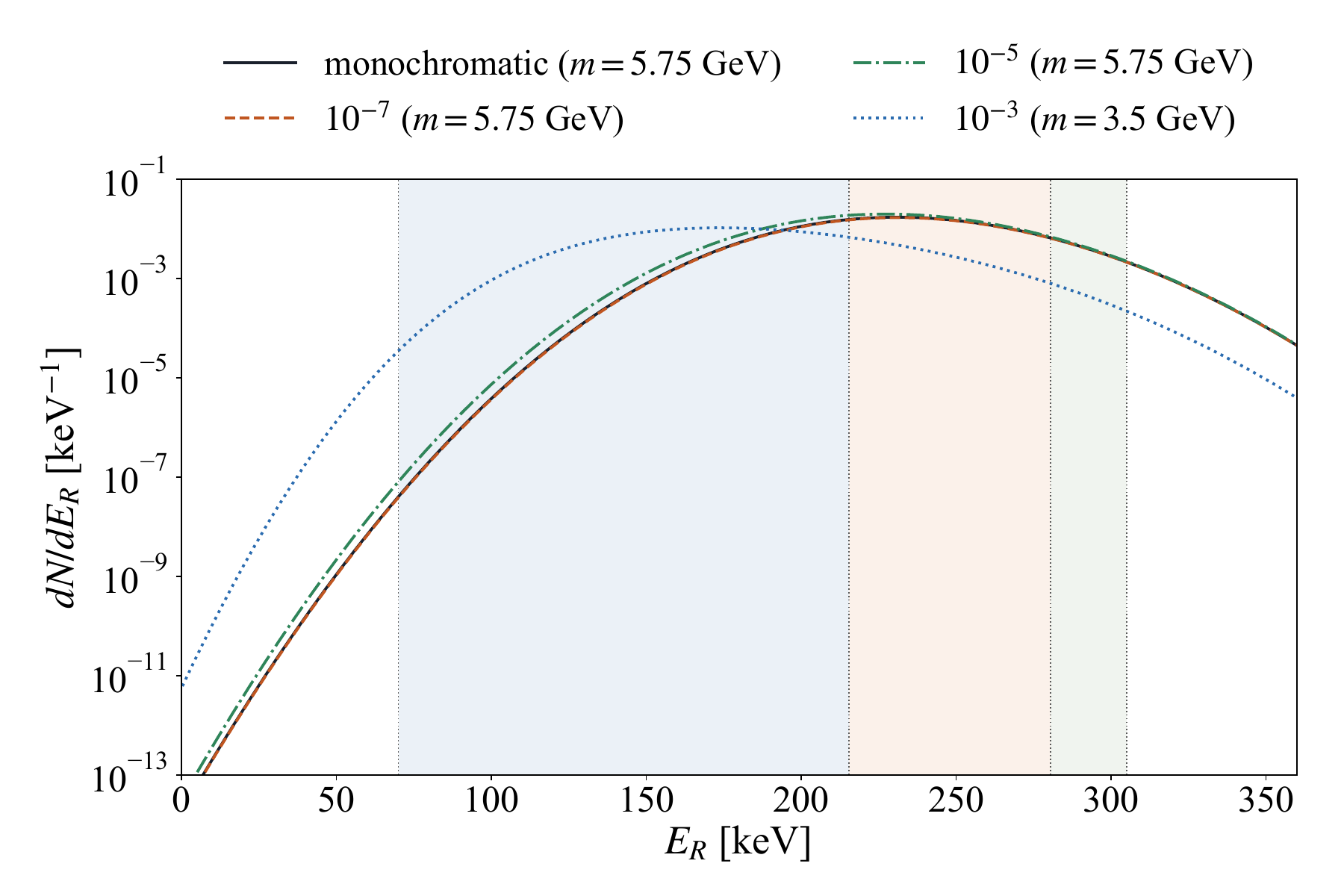}
    \caption{Recoil-energy spectra for different incident energy distributions at $\Phi_0=10^5\,\mathrm{cm}^{-2}\mathrm{s}^{-1}$ and $v_0/c=0.05$. Each curve is labeled with its benchmark mass $m$. The solid curve denotes the monochromatic case. Other line styles indicate Gaussian incident energy distributions with $\sigma_{\rm in}/E_c=10^{-7}$, $10^{-5}$, and $10^{-3}$.}
    \label{fig_sigmaE}
\end{figure}

For $v_0=0.05\,c$ and $\Phi_0=10^5\,\mathrm{cm}^{-2}\mathrm{s}^{-1}$, we study the effect of the Gaussian width $\sigma_{\rm in}$ of the incident energy distribution. FIG.~\ref{fig_sigmaE} shows the recoil spectra for different widths and gives the corresponding best-fit mass $m$ for each case. 
Table~\ref{tab:onset_lzE} gives the minimum mass $m_{\rm min}$
for each speed and width. Below this mass, the upper end of
the incident energy range used in the calculation lies below
the ${}^{136}\mathrm{Xe}$ scattering threshold, so the calculated
rate vanishes. A broader Gaussian lowers $m_{\min}$ because its high-energy tail can reach the threshold even when the central energy $E_c$ lies below it. At fixed $v_0$ and fractional
width $\sigma_{\rm in}/E_c$, the absolute width grows with $m$.
The lower-speed cases start at larger $m_{\rm min}$ in the
monochromatic limit and show a larger drop in $m_{\rm min}$ as
the $\sigma_{\rm in}$ increases. The minimum masses for different
speeds thus move closer together.
As shown in FIG.~\ref{fig_sigmaE}, the maximum of the recoil
spectrum shifts to lower $E_R$ as the best-fit incident mass decreases.
\begin{table}[t]
\centering
\setlength{\heavyrulewidth}{0.8pt}
\setlength{\lightrulewidth}{0.5pt}
\setlength{\cmidrulewidth}{0.5pt}
\renewcommand{\arraystretch}{1.35}   
\caption{Lower edge $m_{\min}$ of the mass scan for each incident
speed in the monochromatic case and for Gaussian widths
$\sigma_{\rm in}/E_c=10^{-7},\,10^{-5},\,10^{-3}$. Below $m_{\min}$ the upper end of the
initial-energy distribution stays below the $^{136}$Xe
threshold and the rate vanishes.}
\label{tab:onset_lzE}
\vspace{4pt}
\begin{tabular}{c c c}
\toprule[0.8pt]
\multirow{2}{*}{$v_0\,[c]$} & \multicolumn{2}{c}{$m_{\min}$ [GeV]} \\
\cmidrule[0.5pt](lr){2-3}
 & monochromatic & $\sigma_{\rm in}=10^{-7}E_c$ \\
\midrule[0.5pt]
0.005 & 72.72 & 71.54 \\
0.01 & 29.96 & 29.86 \\
0.05 & 5.104 & 5.103 \\
0.1 & 2.501 & 2.501 \\
\bottomrule[0.8pt]
\end{tabular}
\vspace{1.2ex}
\begin{tabular}{c c c}
\toprule[0.8pt]
\multirow{2}{*}{$v_0\,[c]$} & \multicolumn{2}{c}{$m_{\min}$ [GeV]} \\
\cmidrule[0.5pt](lr){2-3}
 & $\sigma_{\rm in}=10^{-5}E_c$ & $\sigma_{\rm in}=10^{-3}E_c$ \\
\midrule[0.5pt]
0.005 & 33.05 & 3.241 \\
0.01 & 22.72 & 3.22 \\
0.05 & 5.041 & 2.717 \\
0.1 & 2.493 & 1.969 \\
\bottomrule[0.8pt]
\end{tabular}
\end{table}

The benchmark flux, $\Phi_0=10^{5}\ \mathrm{cm^{-2}s^{-1}}$, is obtained at the largest portal coupling compatible with the constraints of Sec.~\ref{Sec:result}. Since $N_i\propto\Phi_0G_{a\gamma\gamma'}^2$, it constitutes a lower bound on the flux that any source must provide for this $(m,\delta)$. Solar reflection~\cite{Emken:2021lgc} does not meet this requirement. The attainable boost is bounded by the solar escape speed, $v\lesssim5\times10^{-3}c$, an order of magnitude below $v_0=0.05c$; the corresponding elastic recoils in the Sun reach at most tens of keV for GeV-scale DM, two to three orders of magnitude below the $7.2$~MeV
kinetic energy required at $m=5.74$~GeV.

For a standard NFW profile and a thermal cross section, the corresponding flux is
$\simeq5\times10^{-6}\ \mathrm{cm^{-2}s^{-1}}$~\cite{Agashe:2014yua}, and reaching $10^{5}$ would
require $\langle\sigma v\rangle\sim10^{-15}\ \mathrm{cm^{3}s^{-1}}$. For
annihilation channels containing $e^{+}e^{-}$ or photons, this far exceeds the
Planck constraint $\langle\sigma v\rangle\lesssim10^{-26}\ \mathrm{cm^{3}s^{-1}}$~\cite{Slatyer:2015jla,Kawasaki:2021etm}. Capture and
annihilation in the Sun yield a yet smaller
flux~\cite{Fornal:2020npv}. Astrophysical accelerator such as a blazar jet can accelerate dark matter to this velocity
~\cite{Wang:2021jic}, and its predicted flux is model-dependent~\cite{Barillier:2025xct}. We therefore interpret the benchmark as
a requirement on the source rather than as a realized scenario.

\FloatBarrier

\section{Conclusion}
\label{Sec:conclude}
The LZ experiment extended its nuclear-recoil search to approximately $270\,\mathrm{keV}$ and reported a candidate at $248\pm23\,\mathrm{(stat)}\pm23\,\mathrm{(sys)}\,\mathrm{keV}$ in an exposure of $2.84\,\mathrm{tonne\,yr}$. We have investigated the photon-mediated endothermic transition of ALP and dark photon $aA\rightarrow \gamma'A$ to interpret this signal. The analysis has been explicitly done by including natural xenon isotopes, nuclear charge form factors, detector efficiency, as well as energy resolution both in case of non-relativistic and boosted DM scenarios. We compared the reconstructed event counts with the assigned LZ counts using a simplified three-bin Poisson likelihood. The benchmarks are selected by minimizing the corresponding deviance and predict approximately one event in total, with the signal dominated by the middle bin. 

For non-relativistic halo DM, we have calculated the event-numbers by integrating over a fixed laboratory-frame SHM distribution. At the benchmark point, we have found the best-fit point with a mass splitting $\delta\simeq346~{\rm keV}$ at a heavy mass of $m\simeq1.2~{\rm TeV}$ with a coupling of $G_{a\gamma\gamma'}\sim 10^{-1}{\rm GeV}^{-1}$ for one-event producing. This value locates at the adopted upper coupling bound. About $61\%$ of the signal falls in the middle reconstructed-energy bin. The signal comes from the high-speed tail of SHM and is dominated by ${}^{132}\mathrm{Xe}$, ${}^{134}\mathrm{Xe}$, and ${}^{136}\mathrm{Xe}$. For boosted DM, we have treated the initial flux as input and evaluated the monochromatic and Gaussian distributing incident DM. According to the experimental and astrophysical constraint, a benchmark with $m\simeq5.74\,\mathrm{GeV}$ and $\delta\simeq6.66\,\mathrm{MeV}$ provides a comparable fit for $v_0=0.05c$ and $\Phi_0=10^{5}\,\mathrm{cm^{-2}\,s^{-1}}$ in case of mono-energetic DM. It predicts that the middle bin contains approximately $62\%$ of the total signal. A Gaussian energy spread allows boosted DM to scatter even when the central energy lies below threshold.
If the energy spread is sufficiently large, the mass thresholds for different incident speeds can lie very close to one another.

Both scenarios can explain the LZ candidate in the three-bin analysis, with lighter DM masses allowed in the boosted case. For halo DM, the sparse high-speed tail and threshold-suppressed cross section require the coupling to approach its upper bound. The theoretical status of this benchmark remains an open question, although experimental constraints at high masses are weak. The boosted benchmark, by contrast, requires an incident flux best read as a requirement on a non-thermal source rather than as a realized scenario.

\section{acknowledgments}
TL is supported in part by the National Key Research and Development Program of China Grant No. 2020YFC2201504, by the Projects No. 11875062, No. 11947302, No. 12047503, and No. 12275333 supported by the National Natural Science Foundation of China, by the Key Research Program of the Chinese Academy of Sciences, Grant No. XDPB15, by the Scientific Instrument Developing Project of the Chinese Academy of Sciences, Grant No. YJKYYQ20190049, by the International Partnership Program of Chinese Academy of Sciences for Grand Challenges, Grant No. 112311KYSB20210012, and by the Henan Province Outstanding Foreign Scientist Studio Project, No.GZS2025008. This work was supported by the High Performance Computing Platform of Henan Normal University. 

\appendix

\section{Derivation of the cross section}\label{app:amplitude}
\subsection{Vertex and amplitude}
The relevant interactions are
\begin{equation}
\mathcal{L}_{\mathrm{int}}
=
\frac{G_{a\gamma\gamma^{\prime}}}{2}
aF_{\mu\nu}\widetilde F^{\prime\mu\nu}
-eA_\mu J_{\mathrm{em}}^\mu,
\label{eq:app_interaction}
\end{equation}
where $J_{\mathrm{em}}^\mu$ is the electromagnetic
current with the elementary charge $e$ factored out. The coherent scalar-charge current, with the charge form factor separated, is
\begin{equation}
\langle A(k')|J_{\rm em}^{\nu}|A(k)\rangle
=ZF_A(Q)K^\nu,\quad K^{\nu}=k^{\nu}+k'^{\nu},
\label{eq:current}
\end{equation}
where $Q=|\boldsymbol q|$ and $F_A(Q)$ is the
charge form factor used in the main text.
This is the elastic electromagnetic current of a
spin-zero nucleus. For spinful isotopes, we retain
the coherent charge contribution represented by
this current. Nuclear magnetic responses are
outside the calculation. The portal vertex and photon propagator give, up to an overall phase,
\begin{equation}
\mathcal M=
\frac{eG_{a\gamma\gamma'}ZF_A(t)}{t}K_\mu
T^{\mu\nu}
\varepsilon_\nu^*(p').
\label{eq:amplitude}
\end{equation}
Here $\varepsilon_\nu(p^{\prime})$ is the
outgoing dark-photon polarization vector. Here
\begin{equation}
T^{\mu\nu}=\epsilon^{\mu\nu\rho\sigma}q_{\rho}p'_{\sigma}.
\end{equation}
The antisymmetric vertex satisfies
\begin{equation}
q_\mu T^{\mu\nu}=0,
\qquad
p_\nu^{\prime}T^{\mu\nu}=0.
\label{eq:app_vertex_transversality}
\end{equation}

The squared amplitude with dark photon helicities summed is
\begin{equation}
\overline{|\mathcal{M}|}^2\equiv\sum_{\lambda}|\mathcal{M}|^2 =\bigg(\frac{eG_{a\gamma\gamma'}ZF_{A}(t)}{t}\bigg)^2 (-A_{\sigma}A^{\sigma}).
\end{equation}
Here, the vector $A_{\sigma}$ is
\begin{equation}
A_{\sigma}=K^{\nu}T_{\nu\sigma},
\end{equation}
here, the relation
\begin{equation}\label{eq:Dphoton_sum}
\sum_{\lambda}\epsilon'^{\mu}_{\lambda}(p')\epsilon'^{*\nu}_{\lambda}(p')=-g^{\mu\nu}+\frac{p'^{\mu}p'^{\nu}}{m^{2}_{\gamma'}},
\end{equation}
has been used, and the second term's contribution vanishes due to Eq.~\eqref{eq:app_vertex_transversality}.

To evaluate the contraction, introduce the
Mandelstam variables
$s=(p+k)^2$ and
$u=(p-k^{\prime})^2$, which satisfy
$s+t+u
=m^2+m_{\gamma^{\prime}}^2+2M_A^2$. Therefore, one has
\begin{equation}\label{eq:sqM_axion}
\begin{aligned}
\overline{|\mathcal{M}|}^2 =& \frac{Z^2e^2G_{a\gamma\gamma'}^2}{t^2}F^2_{A}(t)
\\
&\times\bigg[Q^2\bigg(tm^2_{X}-(q\cdot k)^2\bigg)-t(Q\cdot k)^2\bigg]
\\
=& -{1\over4}\frac{Z^2e^2G_{a\gamma\gamma'}^2}{t^2}F^2_{A}(t)
\\
&\times\bigg[Q^2\lambda(t,m^2_{\gamma'},m^2_{a})+t(s-u)^2\bigg],
\end{aligned}
\end{equation}
where $Q^2=4m^2_{A}-t$ and $\lambda$ is the Källén function
\begin{equation}
\lambda(a,b,c)=a^2+b^2+c^2-2ab-2bc-2ac.
\end{equation}
No expansion in the incident speed, recoil energy,
or mass splitting has been used in
Eq.~\eqref{eq:sqM_axion}. Alternatively, the squared amplitude can also be written in another form
\begin{equation}
\sum_{\lambda}|\mathcal{M}|^2
=
\frac{
4e^2G_{a\gamma\gamma^{\prime}}^{\,2}
Z^2M_A^2F_A^2(Q)}
{t^2}
\mathcal{K}_A,
\label{eq:app_lab_squared_amplitude}
\end{equation}
where $\mathcal{K}_A=|\boldsymbol p\times\boldsymbol q|^2$. The exact on-shell relations are
\begin{align}
Q^2&=2M_AE_R+E_R^2,
\nonumber\\
\boldsymbol p\cdot\boldsymbol q
&=(E_a+M_A)E_R+m\delta+\frac{\delta^2}{2}.
\label{eq:app_lab_identities}
\end{align}
The second relation follows from
$(p-q)^2=(m+\delta)^2$ and $q^0=E_R$.
It involves a three-vector scalar product,
distinct from the four-vector product
$q\cdot p^{\prime}$ used above.
Therefore,
\begin{align}
\mathcal{K}_A
&=
P^2Q^2-(\boldsymbol p\cdot\boldsymbol q)^2
\nonumber\\
&=
P^2(2M_AE_R+E_R^2)
-
\left[
(E_a+M_A)E_R+m\delta+\frac{\delta^2}{2}
\right]^2.
\label{eq:app_exact_kinematic_factor}
\end{align}

\subsection{Non-relativistic limit}
In this section, the non-relativistic form of the cross section is shown. The general two-body phase-space relation is
\begin{equation}
\frac{\mathrm{d}\sigma_A}{\mathrm{d}t}
=
\frac{\overline{|\mathcal{M}|}^2}
{16\pi\lambda_i}.
\label{eq:app_phase_space}
\end{equation}
For a stationary target,
$k^\mu=(M_A,\boldsymbol 0)$ and
$t=-2M_AE_R$, where
$E_R=k^{\prime0}-M_A$.
It follows that
\begin{equation}
\frac{\mathrm{d}\sigma_A}{\mathrm{d}E_R}
=
\frac{M_A}{8\pi\lambda_i}
\overline{|\mathcal{M}|}^2.
\label{eq:app_recoil_phase_space}
\end{equation}
Substituting Eq.~\eqref{eq:app_lab_squared_amplitude} into
Eq.~\eqref{eq:app_recoil_phase_space} gives
\begin{equation}
\frac{\mathrm{d}\sigma_A}
{\mathrm{d}E_R}
=
\frac{
\alpha G_{a\gamma\gamma^{\prime}}^{\,2}
Z^2F_A^2(Q)}
{8M_AP^2E_R^2}
\mathcal{K}_A,
\label{eq:app_relativistic_recoil_cross_section}
\end{equation}
where $\alpha=e^2/(4\pi)$.

This expression applies on the physical recoil
interval, where $\mathcal{K}_A\geq0$ and
$E_a-E_R\geq m+\delta$.
The reaction requires
$s\geq(M_A+m+\delta)^2$. 

We now take $v\ll1$, $E_R\ll M_A$, and $\delta\ll m$,
retaining $\delta$ at the same order as the kinetic
and recoil energies.
At leading order,
$P\simeq mv$, $E_a\simeq m+mv^2/2$, and
$Q^2\simeq2M_AE_R$.
Equation~\eqref{eq:app_lab_identities} becomes
\begin{align}
\boldsymbol p\cdot\boldsymbol q
&\simeq
(m+M_A)E_R+m\delta
=
m\left(\frac{M_AE_R}{\mu_A}+\delta\right),
\label{eq:app_nr_scalar_product}
\end{align}
where $\mu_A=mM_A/(m+M_A)$ is the ALP--nucleus
reduced mass. Its full mass dependence is retained.

Let $\theta$ be the angle between
$\boldsymbol p$ and $\boldsymbol q$.
Equation~\eqref{eq:app_nr_scalar_product} gives
\begin{equation}
\cos\theta
\simeq
\frac{M_AE_R/\mu_A+\delta}
{v\sqrt{2M_AE_R}}
=
\frac{v_{\min,A}(E_R)}{v},
\label{eq:app_nr_angle}
\end{equation}
where
\begin{equation}
v_{\min,A}(E_R)
=
\frac{M_AE_R/\mu_A+\delta}
{\sqrt{2M_AE_R}}.
\label{eq:app_minimum_speed}
\end{equation}
For $\delta>0$, the condition $|\cos\theta|\leq1$
requires $v\geq v_{\min,A}(E_R)$.

The kinematic factor consequently reduces to
\begin{align}
\mathcal{K}_A
&=
P^2Q^2(1-\cos^2\theta)
\nonumber\\
&\simeq
2M_AP^2E_R
\left[
1-\frac{v_{\min,A}^2(E_R)}{v^2}
\right].
\label{eq:app_nr_kinematic_factor}
\end{align}
Substitution into
Eq.~\eqref{eq:app_relativistic_recoil_cross_section}
yields
\begin{align}
\frac{\mathrm{d}\sigma_A}{\mathrm{d}E_R}
={}&
\frac{
\alpha G_{a\gamma\gamma^{\prime}}^{\,2}Z^2}
{4E_R}
F_A^2(Q)
\nonumber\\
&\times
\left[
1-\frac{v_{\min,A}^2(E_R)}{v^2}
\right]
\Theta\!\left(v-v_{\min,A}(E_R)\right),
\label{eq:app_nr_cross_section}
\end{align}
with $Q\simeq\sqrt{2M_AE_R}$.
The Heaviside function $\Theta$ restricts the
cross section to kinematically accessible recoils.
Equation~\eqref{eq:app_nr_cross_section} reproduces
Eq.~\eqref{eq:differential_cross_section}
in the main text.

At either recoil endpoint,
$v_{\min,A}(E_R)=v$ and
$\boldsymbol p\times\boldsymbol q=0$.
The vanishing cross section at these endpoints
therefore follows from the transverse-momentum
structure of the portal vertex.

\clearpage
\bibliography{dmref}

@article{Rubin:1970zza,
    author = "Rubin, Vera C. and Ford, Jr., W. Kent",
    title = "{Rotation of the Andromeda Nebula from a Spectroscopic Survey of Emission Regions}",
    journal = "Astrophys. J.",
    volume = "159",
    pages = "379--403",
    year = "1970"
}

@article{Clowe:2006eq,
    author = "Clowe, Douglas and Bradac, Marusa and Gonzalez, Anthony H. and Markevitch, Maxim and Randall, Scott W. and Jones, Christine and Zaritsky, Dennis",
    title = "{A direct empirical proof of the existence of dark matter}",
    eprint = "astro-ph/0608407",
    archivePrefix = "arXiv",
    reportNumber = "SLAC-PUB-12078",
    journal = "Astrophys. J. Lett.",
    volume = "648",
    pages = "L109--L113",
    year = "2006"
}

@article{Planck:2018vyg,
    author = "Aghanim, N. and others",
    collaboration = "Planck",
    title = "{Planck 2018 results. VI. Cosmological parameters}",
    eprint = "1807.06209",
    archivePrefix = "arXiv",
    primaryClass = "astro-ph.CO",
    journal = "Astron. Astrophys.",
    volume = "641",
    pages = "A6",
    year = "2020",
    note = "[Erratum: Astron.Astrophys. 652, C4 (2021)]"
}

@article{Essig:2015cda,
    author = "Essig, Rouven and Fernandez-Serra, Marivi and Mardon, Jeremy and Soto, Adrian and Volansky, Tomer and Yu, Tien-Tien",
    title = "{Direct Detection of sub-GeV Dark Matter with Semiconductor Targets}",
    eprint = "1509.01598",
    archivePrefix = "arXiv",
    primaryClass = "hep-ph",
    journal = "JHEP",
    volume = "05",
    pages = "046",
    year = "2016"
}

@article{Essig:2017kqs,
    author = "Essig, Rouven and Volansky, Tomer and Yu, Tien-Tien",
    title = "{New Constraints and Prospects for sub-GeV Dark Matter Scattering off Electrons in Xenon}",
    eprint = "1703.00910",
    archivePrefix = "arXiv",
    primaryClass = "hep-ph",
    reportNumber = "CERN-TH-2017-042, YITP-SB-17-09",
    journal = "Phys. Rev. D",
    volume = "96",
    number = "4",
    pages = "043017",
    year = "2017"
}

@article{Crisler:2018gci,
    author = "Crisler, Michael and Essig, Rouven and Estrada, Juan and Fernandez, Guillermo and Tiffenberg, Javier and Sofo haro, Miguel and Volansky, Tomer and Yu, Tien-Tien",
    collaboration = "SENSEI",
    title = "{SENSEI: First Direct-Detection Constraints on sub-GeV Dark Matter from a Surface Run}",
    eprint = "1804.00088",
    archivePrefix = "arXiv",
    primaryClass = "hep-ex",
    reportNumber = "FERMILAB-PUB-18-116-AE-PPD, YITP-SB-18-4, CERN-TH-2018-070",
    journal = "Phys. Rev. Lett.",
    volume = "121",
    number = "6",
    pages = "061803",
    year = "2018"
}

@article{SENSEI:2023zdf,
    author = "Adari, Prakruth and others",
    collaboration = "SENSEI",
    title = "{First Direct-Detection Results on Sub-GeV Dark Matter Using the SENSEI Detector at SNOLAB}",
    eprint = "2312.13342",
    archivePrefix = "arXiv",
    primaryClass = "astro-ph.CO",
    reportNumber = "YITP-SB-2023-30, FERMILAB-PUB-23-0824-CSAID-PPD",
    doi = "10.1103/PhysRevLett.134.011804",
    journal = "Phys. Rev. Lett.",
    volume = "134",
    number = "1",
    pages = "011804",
    year = "2025"
}

@article{Lewin:1995rx,
    author = "Lewin, J. D. and Smith, P. F.",
    title = "{Review of mathematics, numerical factors, and corrections for dark matter experiments based on elastic nuclear recoil}",
    reportNumber = "RAL-TR-95-024",
    journal = "Astropart. Phys.",
    volume = "6",
    pages = "87--112",
    year = "1996"
}

@article{LZ:2022lsv,
    author = "Aalbers, J. and others",
    collaboration = "LZ",
    title = "{First Dark Matter Search Results from the LUX-ZEPLIN (LZ) Experiment}",
    eprint = "2207.03764",
    archivePrefix = "arXiv",
    primaryClass = "hep-ex",
    journal = "Phys. Rev. Lett.",
    volume = "131",
    number = "4",
    pages = "041002",
    year = "2023"
}

@article{XENON:2023cxc,
    author = "Aprile, E. and others",
    collaboration = "XENON",
    title = "{First Dark Matter Search with Nuclear Recoils from the XENONnT Experiment}",
    eprint = "2303.14729",
    archivePrefix = "arXiv",
    primaryClass = "hep-ex",
    journal = "Phys. Rev. Lett.",
    volume = "131",
    number = "4",
    pages = "041003",
    year = "2023"
}

@article{PandaX:2022aac,
    author = "Ma, Wenbo and others",
    collaboration = "PandaX",
    title = "{Search for Solar B8 Neutrinos in the PandaX-4T Experiment Using Neutrino-Nucleus Coherent Scattering}",
    eprint = "2207.04883",
    archivePrefix = "arXiv",
    primaryClass = "hep-ex",
    journal = "Phys. Rev. Lett.",
    volume = "130",
    number = "2",
    pages = "021802",
    year = "2023"
}

@article{SuperCDMS:2024yiv,
    author = "Albakry, M. F. and others",
    collaboration = "SuperCDMS",
    title = "{Light dark matter constraints from SuperCDMS HVeV detectors operated underground with an anticoincidence event selection}",
    eprint = "2407.08085",
    archivePrefix = "arXiv",
    primaryClass = "hep-ex",
    reportNumber = "FERMILAB-PUB-24-0376-PPD",
    journal = "Phys. Rev. D",
    volume = "111",
    number = "1",
    pages = "012006",
    year = "2025"
}

@article{LZ:2026axp,
    author = "Akerib, D. S. and others",
    collaboration = "LZ",
    title = "{Search for dark matter particle interactions in an extended nuclear recoil energy window with the LUX-ZEPLIN (LZ) experiment}",
    eprint = "2609.02823",
    archivePrefix = "arXiv",
    primaryClass = "hep-ex",
    month = "9",
    year = "2026"
}

@article{PDG2025,
  author  = {Navas, S. and others},
  title   = {Review of Particle Physics},
  journal = {Physical Review D},
  volume  = {110},
  pages   = {030001},
  year    = {2024},
  note    = {2025 update, Sec. 27.6}
}

@article{Graham:2010ca,
    author = "Graham, Peter W. and Harnik, Roni and Rajendran, Surjeet and Saraswat, Prashant",
    title = "{Exothermic Dark Matter}",
    eprint = "1004.0937",
    archivePrefix = "arXiv",
    primaryClass = "hep-ph",
    reportNumber = "FERMILAB-PUB-10-062-T, MIT-CTP-4140, SU-ITP-10-15",
    journal = "Phys. Rev. D",
    volume = "82",
    pages = "063512",
    year = "2010"
}

@article{Bell:2021zkr,
    author = "Bell, Nicole F. and Dent, James B. and Dutta, Bhaskar and Ghosh, Sumit and Kumar, Jason and Newstead, Jayden L.",
    title = "{Low-mass inelastic dark matter direct detection via the Migdal effect}",
    eprint = "2103.05890",
    archivePrefix = "arXiv",
    primaryClass = "hep-ph",
    reportNumber = "MI-TH-215",
    journal = "Phys. Rev. D",
    volume = "104",
    number = "7",
    pages = "076013",
    year = "2021"
}

@article{Tucker-Smith:2001myb,
    author = "Tucker-Smith, David and Weiner, Neal",
    title = "{Inelastic dark matter}",
    eprint = "hep-ph/0101138",
    archivePrefix = "arXiv",
    reportNumber = "UCB-PTH-00-43, LBNL-47234, UW-PT-00-17",
    journal = "Phys. Rev. D",
    volume = "64",
    pages = "043502",
    year = "2001"
}

@article{XENON:2022avm,
    author = "Aprile, E. and others",
    collaboration = "XENON",
    title = "{Effective field theory and inelastic dark matter results from XENON1T}",
    eprint = "2210.07591",
    archivePrefix = "arXiv",
    primaryClass = "hep-ex",
    journal = "Phys. Rev. D",
    volume = "109",
    number = "11",
    pages = "112017",
    year = "2024"
}

@article{Jaeckel:2010ni,
    author = "Jaeckel, Joerg and Ringwald, Andreas",
    title = "{The Low-Energy Frontier of Particle Physics}",
    eprint = "1002.0329",
    archivePrefix = "arXiv",
    primaryClass = "hep-ph",
    reportNumber = "CPT-10-18, DESY-10-016, IPPP-10-09",
    journal = "Ann. Rev. Nucl. Part. Sci.",
    volume = "60",
    pages = "405--437",
    year = "2010"
}

@inproceedings{Ringwald:2014vqa,
    author = "Ringwald, A.",
    title = "{Axions and Axion-Like Particles}",
    booktitle = "{49th Rencontres de Moriond on Electroweak Interactions and Unified Theories}",
    eprint = "1407.0546",
    archivePrefix = "arXiv",
    primaryClass = "hep-ph",
    reportNumber = "DESY-14-108",
    pages = "223--230",
    year = "2014"
}

@article{Holdom:1985ag,
    author = "Holdom, Bob",
    title = "{Two U(1)'s and Epsilon Charge Shifts}",
    reportNumber = "UTPT-85-30",
    journal = "Phys. Lett. B",
    volume = "166",
    pages = "196--198",
    year = "1986"
}

@article{He:2020sat,
    author = "He, Hong-Jian and Wang, Yu-Chen and Zheng, Jiaming",
    title = "{GeV-scale inelastic dark matter with dark photon mediator via direct detection and cosmological and laboratory constraints}",
    eprint = "2012.05891",
    archivePrefix = "arXiv",
    primaryClass = "hep-ph",
    journal = "Phys. Rev. D",
    volume = "104",
    number = "11",
    pages = "115033",
    year = "2021"
}

@article{PandaX:2023xgl,
    author = "Huang, Di and others",
    collaboration = "PandaX",
    title = "{Search for Dark-Matter{\textendash}Nucleon Interactions with a Dark Mediator in PandaX-4T}",
    eprint = "2308.01540",
    archivePrefix = "arXiv",
    primaryClass = "hep-ex",
    journal = "Phys. Rev. Lett.",
    volume = "131",
    number = "19",
    pages = "191002",
    year = "2023"
}

@article{Bae:2023ago,
    author = "Bae, Kyu Jung and Kim, Jongkuk",
    title = "{Axion-Mediated Inelastic Dark Matter}",
    eprint = "2312.11210",
    archivePrefix = "arXiv",
    primaryClass = "hep-ph",
    month = "12",
    year = "2023"
}

@article{Beenakker:2025mhf,
    author = {Beenakker, Wim and Mikkers, Dani{\"e}l and Phan, Anh Vu and Westhoff, Susanne},
    title = "{ALP-mediated dark matter-nucleon scattering}",
    eprint = "2511.19619",
    archivePrefix = "arXiv",
    primaryClass = "hep-ph",
    reportNumber = "Nikhef 2025-017",
    journal = "JHEP",
    volume = "05",
    pages = "033",
    year = "2026"
}

@article{Yuan:2026djt,
    author = "Yuan, Guan-Wen and Zhang, Bo and Cao, Wen-Yu and Feng, Lei and Yang, Ruizhi",
    title = "{ALP-mediated inelastic dark matter and the LUX-ZEPLIN high-recoil candidate event LZ230616}",
    eprint = "2609.08893",
    archivePrefix = "arXiv",
    primaryClass = "hep-ph",
    month = "9",
    year = "2026"
}

@article{Unwin:2026rdp,
    author = "Unwin, James",
    title = "{Axion Portal Dark Matter and the LUX-ZEPLIN High-Recoil Event}",
    eprint = "2609.04186",
    archivePrefix = "arXiv",
    primaryClass = "hep-ph",
    month = "9",
    year = "2026"
}

@article{Yamashita:2026ump,
    author = "Yamashita, Kimiko",
    title = "{Inelastic Dark Photon Dark Matter for the LUX-ZEPLIN High-Recoil Event and the Galactic Halo Gamma-Ray Excess}",
    eprint = "2609.02868",
    archivePrefix = "arXiv",
    primaryClass = "hep-ph",
    month = "9",
    year = "2026"
}

@article{deLima:2026shq,
    author = "de Lima, Carlos Henrique",
    title = "{Exothermic Dark Matter at LZ}",
    eprint = "2609.05204",
    archivePrefix = "arXiv",
    primaryClass = "hep-ph",
    month = "9",
    year = "2026"
}

@article{Zhu:2026dag,
    author = "Zhu, Pengxuan and Dalla Valle Garcia, Giovani and Wang, Xuan-Gong and Thomas, Anthony W. and White, Martin J.",
    title = "{Endothermic dark matter with a light dark photon and the LUX-ZEPLIN high-energy nuclear-recoil candidate}",
    eprint = "2609.09015",
    archivePrefix = "arXiv",
    primaryClass = "hep-ph",
    reportNumber = "ADP-26-15/T1312",
    month = "9",
    year = "2026"
}

@article{Yin:2026jnn,
    author = "Yin, Wen",
    title = "{A PQ-Symmetric High-Scale SUSY Interpretation of the LZ High-Energy Recoil}",
    eprint = "2609.01892",
    archivePrefix = "arXiv",
    primaryClass = "hep-ph",
    month = "9",
    year = "2026"
}

@article{Fan:2026kxx,
    author = "Fan, JiJi and Reece, Matthew",
    title = "{Higgsino Above the Sea of Fog}",
    eprint = "2609.01504",
    archivePrefix = "arXiv",
    primaryClass = "hep-ph",
    month = "9",
    year = "2026"
}

@article{Freese:2026sga,
    author = "Freese, Katherine and Theodosopoulos, Dionysios P.",
    title = "{Higgsino Dark Matter Interpretation of the LUX-ZEPLIN 248 keV Nuclear-Recoil Event}",
    eprint = "2609.01583",
    archivePrefix = "arXiv",
    primaryClass = "hep-ph",
    month = "9",
    year = "2026"
}

@article{Du:2026guj,
    author = "Du, Xiaokang and Wang, Fei",
    title = "{TeV Higgsino Interpretation of the LZ High-Recoil Event with Intermediate-Scale Electroweak Gauginos}",
    eprint = "2609.04163",
    archivePrefix = "arXiv",
    primaryClass = "hep-ph",
    month = "9",
    year = "2026"
}

@article{Lian:2026hpm,
    author = "Lian, Jingwei and Yang, Jin Min",
    title = "{Explaining the LZ High-Energy Recoil Event with Inelastic Sneutrino Dark Matter in Supersymmetry}",
    eprint = "2609.15742",
    archivePrefix = "arXiv",
    primaryClass = "hep-ph",
    month = "9",
    year = "2026"
}

@article{Roy:2026wzb,
    author = "Roy, Subhojit and Schwaller, Pedro and Wagner, Carlos E. M.",
    title = "{Interpreting the High-Recoil LUX-ZEPLIN Event with Bino-/Singlino-like and Higgsino Dark Matter}",
    eprint = "2609.40234",
    archivePrefix = "arXiv",
    primaryClass = "hep-ph",
    month = "9",
    year = "2026"
}

@article{Wang:2026ytg,
    author = "Wang, Lei and Xiao, Yang",
    title = "{The Inert Doublet Model of Dark Matter and the LUX-ZEPLIN High-Recoil Event}",
    eprint = "2609.06571",
    archivePrefix = "arXiv",
    primaryClass = "hep-ph",
    month = "9",
    year = "2026"
}

@article{Bandyopadhyay:2026gjw,
    author = "Bandyopadhyay, Disha and Borah, Debasish and Borah, Pankaj",
    title = "{LZ nuclear recoil event from inelastic singlet-doublet scalar dark matter}",
    eprint = "2609.07451",
    archivePrefix = "arXiv",
    primaryClass = "hep-ph",
    month = "9",
    year = "2026"
}

@article{Borah:2026zwf,
    author = "Borah, Debasish and Sahoo, Sujit Kumar and Sahu, Narendra and Sharma, Shashwat",
    title = "{Inelastic Singlet-Doublet Fermion Dark Matter in light of the 248 keV LZ event}",
    eprint = "2609.07800",
    archivePrefix = "arXiv",
    primaryClass = "hep-ph",
    month = "9",
    year = "2026"
}

@article{Liang:2026coz,
    author = "Liang, Jin-Han and Liu, Zuowei and Tran, Van Que and Xu, Yongheng",
    title = "{LZ Nuclear-Recoil Excess from Boosted Light Magnetic Dipole-dipole Dark Matter}",
    eprint = "2609.06756",
    archivePrefix = "arXiv",
    primaryClass = "hep-ph",
    month = "9",
    year = "2026"
}

@article{Kannike:2026qyl,
    author = "Kannike, Kristjan and Raidal, Martti and Strumia, Alessandro",
    title = "{Boosted dark particles and the LZ nuclear recoil event}",
    eprint = "2609.07742",
    archivePrefix = "arXiv",
    primaryClass = "hep-ph",
    month = "9",
    year = "2026"
}

@article{Alhazmi:2026efz,
    author = "Alhazmi, Haider and Kim, Doojin and Kong, Kyoungchul and Park, Jong-Chul and Shin, Seodong",
    title = "{High-Energy Nuclear Recoils from Boosted Dark Matter for the LZ 248-keV Event: Beyond the Halo-Dependent High-Velocity Tail}",
    eprint = "2609.06890",
    archivePrefix = "arXiv",
    primaryClass = "hep-ph",
    month = "9",
    year = "2026"
}

@article{Heikinheimo:2026kwp,
    author = "Heikinheimo, Matti and Zimmermann, Niklas",
    title = "{Cosmic ray boosted dark matter with momentum dependent interactions can explain the LZ 248 keV event}",
    eprint = "2609.11600",
    archivePrefix = "arXiv",
    primaryClass = "hep-ph",
    month = "9",
    year = "2026"
}

@article{Chauhan:2026udz,
    author = "Chauhan, Bhavesh and Sahasrabuddhe, Soham and Sen, Manibrata",
    title = "{A closer look at the LZ 248 keV event through the lens of cosmic-ray boosted dark matter}",
    eprint = "2609.24982",
    archivePrefix = "arXiv",
    primaryClass = "hep-ph",
    month = "9",
    year = "2026"
}

@article{Kaneta:2016wvf,
    author = "Kaneta, Kunio and Lee, Hye-Sung and Yun, Seokhoon",
    title = "{Portal Connecting Dark Photons and Axions}",
    eprint = "1611.01466",
    archivePrefix = "arXiv",
    primaryClass = "hep-ph",
    reportNumber = "CTPU-16-34",
    journal = "Phys. Rev. Lett.",
    volume = "118",
    number = "10",
    pages = "101802",
    year = "2017"
}

@article{Kaneta:2017wfh,
    author = "Kaneta, Kunio and Lee, Hye-Sung and Yun, Seokhoon",
    title = "{Dark photon relic dark matter production through the dark axion portal}",
    eprint = "1704.07542",
    archivePrefix = "arXiv",
    primaryClass = "hep-ph",
    reportNumber = "CTPU-17-15",
    journal = "Phys. Rev. D",
    volume = "95",
    number = "11",
    pages = "115032",
    year = "2017"
}

@article{Helm:1956zz,
    author = "Helm, Richard H.",
    title = "{Inelastic and Elastic Scattering of 187-Mev Electrons from Selected Even-Even Nuclei}",
    doi = "10.1103/PhysRev.104.1466",
    journal = "Phys. Rev.",
    volume = "104",
    pages = "1466--1475",
    year = "1956"
}

@article{Freese:2012xd,
    author = "Freese, Katherine and Lisanti, Mariangela and Savage, Christopher",
    title = "{Colloquium: Annual modulation of dark matter}",
    eprint = "1209.3339",
    archivePrefix = "arXiv",
    primaryClass = "astro-ph.CO",
    doi = "10.1103/RevModPhys.85.1561",
    journal = "Rev. Mod. Phys.",
    volume = "85",
    pages = "1561--1581",
    year = "2013"
}

@article{Fan:2026hzw,
    author = "Fan, Zi-Tong and He, Hong-Jian and Wang, Yu-Chen and Zhao, Yue",
    title = "{Inelastic Dark Matter and High-Energy Recoil Signatures in LZ}",
    eprint = "2609.10491",
    archivePrefix = "arXiv",
    primaryClass = "hep-ph",
    journal = "Chin. Phys. Lett.",
    volume = "43",
    number = "11",
    pages = "110201",
    year = "2026"
}

@article{Zhang:2023oed,
    author = "Zhang, Wenxing and Li, Tianjun and Yin, Xiangwei",
    title = "{The Z resonance, inelastic dark matter, and new physics anomalies in the Simple Extension of the Standard Model (SESM) with general scalar potential}",
    eprint = "2303.14764",
    archivePrefix = "arXiv",
    primaryClass = "hep-ph",
    journal = "Eur. Phys. J. C",
    volume = "83",
    number = "8",
    pages = "725",
    year = "2023"
}

@article{Agashe:2014yua,
    author = "Agashe, Kaustubh and Cui, Yanou and Necib, Lina and Thaler, Jesse",
    title = "{(In)direct Detection of Boosted Dark Matter}",
    eprint = "1405.7370",
    archivePrefix = "arXiv",
    primaryClass = "hep-ph",
    reportNumber = "MIT-CTP-4538, UMD-PP-014-005",
    journal = "JCAP",
    volume = "10",
    pages = "062",
    year = "2014"
}

@article{Acevedo:2026tur,
    author = "Acevedo, Javier F. and Ritz, Adam",
    title = "{Boosted Dark Matter from Sagittarius A$^\star$}",
    eprint = "2606.30724",
    archivePrefix = "arXiv",
    primaryClass = "hep-ph",
    month = "6",
    year = "2026"
}

@article{Chen:2025hbh,
    author = "Chen, Chuan-Ren and Hsieh, Yuan-Feng and Tran, Van Que",
    title = "{Probing the axion-photon-dark photon interaction at future e+e- colliders}",
    eprint = "2509.08733",
    archivePrefix = "arXiv",
    primaryClass = "hep-ph",
    journal = "Phys. Rev. D",
    volume = "113",
    number = "3",
    pages = "035008",
    year = "2026"
}

@article{Yue:2026oqu,
    author = "Yue, Chong-Xing and Bu, Yang-Yang and Li, Xin-Yang and Tang, Feng-Xia",
    title = "{Probing the ALP-photon-dark photon interaction through the mono-photon plus missing-energy signature at a muon collider}",
    eprint = "2608.15656",
    archivePrefix = "arXiv",
    primaryClass = "hep-ph",
    journal = "Phys. Rev. D",
    volume = "114",
    number = "7",
    pages = "075011",
    year = "2026"
}

@article{Chen:2026beb,
    author = "Chen, Chuan-Ren and Nomura, Takaaki and Tran, Van Que",
    title = "{Probing the Axion--Photon--Dark Photon Interaction with LHC Mono-Photons and Fermi-LAT Gamma-Ray Lines}",
    eprint = "2610.21989",
    archivePrefix = "arXiv",
    primaryClass = "hep-ph",
    month = "10",
    year = "2026"
}

@article{Guha:2018mli,
    author = "Guha, Atanu and Dev, P. S. Bhupal and Das, Prasanta Kumar",
    title = "{Model-independent Astrophysical Constraints on Leptophilic Dark Matter in the Framework of Tsallis Statistics}",
    eprint = "1810.00399",
    archivePrefix = "arXiv",
    primaryClass = "hep-ph",
    doi = "10.1088/1475-7516/2019/02/032",
    journal = "JCAP",
    volume = "02",
    pages = "032",
    year = "2019"
}

@article{Emken:2021lgc,
    author = "Emken, Timon",
    title = "{Solar reflection of light dark matter with heavy mediators}",
    eprint = "2102.12483",
    archivePrefix = "arXiv",
    primaryClass = "hep-ph",
    doi = "10.1103/PhysRevD.105.063020",
    journal = "Phys. Rev. D",
    volume = "105",
    number = "6",
    pages = "063020",
    year = "2022"
}

@article{Slatyer:2015jla,
    author = "Slatyer, Tracy R.",
    title = "{Indirect dark matter signatures in the cosmic dark ages. I. Generalizing the bound on s-wave dark matter annihilation from Planck results}",
    eprint = "1506.03811",
    archivePrefix = "arXiv",
    primaryClass = "hep-ph",
    reportNumber = "MIT-CTP-4682",
    doi = "10.1103/PhysRevD.93.023527",
    journal = "Phys. Rev. D",
    volume = "93",
    number = "2",
    pages = "023527",
    year = "2016"
}

@article{Kawasaki:2021etm,
    author = "Kawasaki, Masahiro and Nakatsuka, Hiromasa and Nakayama, Kazunori and Sekiguchi, Toyokazu",
    title = "{Revisiting CMB constraints on dark matter annihilation}",
    eprint = "2105.08334",
    archivePrefix = "arXiv",
    primaryClass = "astro-ph.CO",
    doi = "10.1088/1475-7516/2021/12/015",
    journal = "JCAP",
    volume = "12",
    number = "12",
    pages = "015",
    year = "2021"
}

@article{Fornal:2020npv,
    author = "Fornal, Bartosz and Sandick, Pearl and Shu, Jing and Su, Meng and Zhao, Yue",
    title = "{Boosted Dark Matter Interpretation of the XENON1T Excess}",
    eprint = "2006.11264",
    archivePrefix = "arXiv",
    primaryClass = "hep-ph",
    doi = "10.1103/PhysRevLett.125.161804",
    journal = "Phys. Rev. Lett.",
    volume = "125",
    number = "16",
    pages = "161804",
    year = "2020"
}

@article{Wang:2021jic,
    author = "Wang, Jin-Wei and Granelli, Alessandro and Ullio, Piero",
    title = "{Direct Detection Constraints on Blazar-Boosted Dark Matter}",
    eprint = "2111.13644",
    archivePrefix = "arXiv",
    primaryClass = "astro-ph.HE",
    doi = "10.1103/PhysRevLett.128.221104",
    journal = "Phys. Rev. Lett.",
    volume = "128",
    number = "22",
    pages = "221104",
    year = "2022"
}

@article{Barillier:2025xct,
    author = "Barillier, Erin and Manenti, Laura and Mora, Knut and Padovani, Paolo and Sarnoff, Isaac and Xu, Yongheng and Penning, Bjorn and Arneodo, Francesco",
    title = "{Setting limits on blazar-boosted dark matter with xenon-based detectors}",
    eprint = "2509.07265",
    archivePrefix = "arXiv",
    primaryClass = "astro-ph.HE",
    doi = "10.1103/qxkb-4bpy",
    journal = "Phys. Rev. D",
    volume = "113",
    number = "2",
    pages = "023005",
    year = "2026"
}

@article{Drukier:1986tm,
    author = "Drukier, A. K. and Freese, Katherine and Spergel, D. N.",
    title = "{Detecting Cold Dark Matter Candidates}",
    doi = "10.1103/PhysRevD.33.3495",
    journal = "Phys. Rev. D",
    volume = "33",
    pages = "3495--3508",
    year = "1986"
}
\bibliographystyle{dm}

\end{document}